\documentclass[manuscript,acmsmall]{acmart}
\AtBeginDocument{%
  \providecommand\BibTeX{{%
    \normalfont B\kern-0.5em{\scshape i\kern-0.25em b}\kern-0.8em\TeX}}}

\setcopyright{acmcopyright}
\copyrightyear{2018}
\acmYear{2018}
\acmDOI{XXXXXXX.XXXXXXX}

\acmConference[Conference acronym 'XX]{Make sure to enter the correct
  conference title from your rights confirmation email}{June 03--05,
  2018}{Woodstock, NY}
\acmPrice{15.00}
\acmISBN{978-1-4503-XXXX-X/18/06}

\usepackage{array}
\usepackage{longtable}
\usepackage{booktabs}
\usepackage{colortbl}
\usepackage{rotating}
\usepackage{float}
\usepackage{fancybox}
\usepackage{color}

\definecolor{greenxian}{RGB}{141,211,199}
\definecolor{yellowxian}{RGB}{255,237,111}
\definecolor{purplexian}{RGB}{190,186,218}
\definecolor{redxian}{RGB}{251,128,114}
\definecolor{bluexian}{RGB}{128,177,211}
\definecolor{orangexian}{RGB}{253,180,98}

\begin{document}

\title[XR-based Remote HRI]{Towards Massive Interaction with Generalist Robotics: A Systematic Review of XR-enabled Remote Human-Robot Interaction Systems}

\author{Xian Wang}
\orcid{0000-0003-1023-636X}
\email{xian0203.wang@polyu.edu.hk}
\affiliation{%
  \institution{The Hong Kong Polytechnic University}
  \city{Hong Kong}
  \country{Hong Kong SAR}
}

\author{Luyao Shen}
\affiliation{%
  \institution{The Hong Kong University of Science and Technology (Guangzhou)}
  \city{Guangzhou}
  \country{China}
}
\email{lshen595@connect.hkust-gz.edu.cn}

\author{Lik-Hang Lee}
\authornote{L.-H. Lee is the corresponding author.}
\affiliation{
  \institution{The Hong Kong Polytechnic University}
  \city{Hong Kong}
  \country{Hong Kong SAR}
}
\email{lik-hang.lee@polyu.edu.hk}


\begin{abstract}
The rising interest of generalist robots seek to create robots with versatility to handle multiple tasks in a variety of environments, and human will interact with such robots through immersive interfaces. In the context of human-robot interaction (HRI), this survey provides an exhaustive review of the applications of extended reality (XR) technologies in the field of remote HRI. We developed a systematic search strategy based on the PRISMA methodology. From the initial 2,561 articles selected, 100 research papers that met our inclusion criteria were included. We categorized and summarized the domain in detail, delving into XR technologies, including augmented reality (AR), virtual reality (VR), and mixed reality (MR), and their applications in facilitating intuitive and effective remote control and interaction with robotic systems. 
The survey highlights existing articles on the application of XR technologies, user experience enhancement, and various interaction designs for XR in remote HRI, providing insights into current trends and future directions. We also identified potential gaps and opportunities for future research to improve remote HRI systems through XR technology to guide and inform future XR and robotics research.
\end{abstract}


\begin{teaserfigure}
  \includegraphics[width=\textwidth]{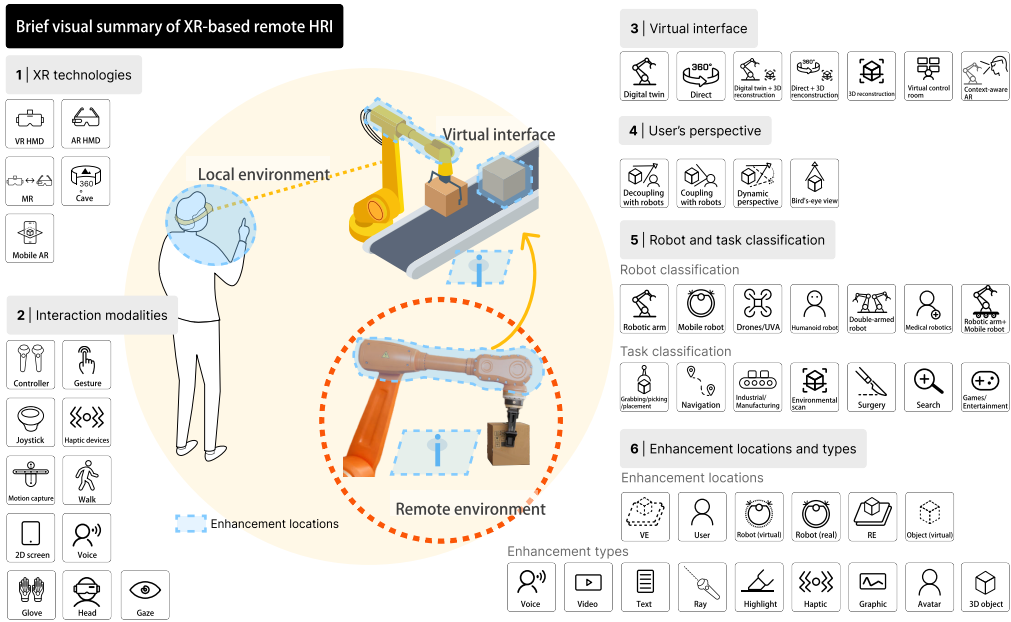}
  \caption{Brief visual summary of the current state of the domain concerning XR-based remote HRI, summarizes the six key dimensions of current system design.}
  \label{fig:teaser}
\end{teaserfigure}

\maketitle

\section{Introduction}
\label{sec:intro}
The field of Human-Robot Interaction (HRI), originally perceived as the study of human engagement with robots, has evolved to investigate the design of robots for socially meaningful interactions with humans and the enhancement of these interactions~\cite{sheridan2016human}. HRI aims to engineer robots capable of effectively collaborating with humans across diverse contexts, such as industrial environments, domestic settings, and educational institutions. In our increasingly globalized and digitalized world, where advances in communication and mobile technologies are driving the rapid development of collaboration concepts, traditional face-to-face interactions are being complemented, and, in certain instances, replaced by remote collaboration mechanisms that overcome geographical and temporal constraints. Remote human-robot collaboration holds considerable potential to revolutionize various fields. For instance, it can facilitate tasks in hazardous environments by allowing humans to control robots from a safe distance, or it can enhance complex tasks by enabling robots to relay information from remote locations to humans. The implications of this form of collaboration include safer workplaces, improved efficiency, and greater accessibility.

Despite videoconferencing and teleconferencing serving as effective tools for remote collaboration, these technologies exhibit limitations when applied to complex tasks within the context of remote human-robot collaboration. Particularly in scenarios involving robotic remote control or teaching, the lack of intuitive and contextual understanding offered by simple video interfaces can pose substantial challenges. To overcome these limitations, Extended Reality (XR) -- encompassing Augmented Reality (AR), Virtual Reality (VR), and Mixed Reality (MR) -- offers a promising solution. AR overlays digital information onto the real world, VR immerses users in a completely digital environment, and MR blends real and virtual worlds. These technologies provide a fusion of the digital and physical worlds, allowing physical and digital objects to coexist and interact in real time. XR technologies are increasingly sophisticated and portable, presenting new opportunities for HRI. Devices such as the Meta Quest 2 and Hololens 2 exemplify the potential of XR to enhance remote HRI. For example, with the assistance of XR technology, novices can perform risky tasks (e.g., welding) in a safe space and in a more intuitive way (e.g., with the first view of the robot). In addition, operators can use XR to switch between different locations to control the robot without needing to move in physical space. XR empowers remote HRI to be more immersive, intuitive, and effective.

However, numerous challenges need to be addressed to unlock XR's full potential in the context of remote human-robot interaction. These challenges include designing intuitive interaction techniques, reducing remote control latency, and evaluating remote HRI systems. Given that this remains an under-researched topic, this paper seeks to contribute to this emerging field of study. We provide a systematic literature review of XR technology-based remote human-robot collaboration applications, highlight key advancements, and identify areas that deserve to be studied in further research. The
contributions of this survey are as follows:

\begin{enumerate}
\item Provide a comprehensive review of HRI system design based on XR technology for remote control scenarios.
\item Identify general convergences and divergences in system design within the existing literature.
\item Propose a research agenda for future XR-based remote HRI.
\end{enumerate}

\subsection{Overview of XR-based Remote Human-Robot Interaction}

Figure \ref{fig:teaser} provides a brief pictorial summary of the current state of the domain concerning XR-based remote HRI. In our proposed model, we establish two distinct spaces: the `\textbf{local space}'
 and the `\textbf{remote space}', 
The survey scope is highly relevant to these two spaces. Thus, we give their definitions as follows.

\subsubsection{Local Space:}

`Local space' refers to the physical environment in which the user is located. It is often equipped with XR technologies, such as AR or VR headsets, that can overlay or immerse the user in a virtual environment (see Fig.\ref{fig:teaser} left of the center schematic). This space is crucial for the user's interaction with the robot, as it hosts the virtual interface for controlling the robot. For example, in the local space, a user wearing a VR headset may see a robotic digital twin (see Fig.\ref{fig:teaser} top right of the center schematic) in the virtual space. This virtual interface allows the user to issue commands to the robot located in another space through various interaction modalities, such as gesture control. The local space is designed to be intuitive and user-friendly, allowing the user to manipulate the robot to perform complex tasks without having to be physically present in the same space as the robot.

\subsubsection{Remote Space:}

Conversely, `remote space' is the physical environment where robots operate and perform their tasks. This could be a factory handling hazardous materials, a distant planet, or a complex surgical field where the direct presence of humans is impossible or undesirable. The robot acts as an agent for the user, performing tasks from the local space via a virtual interface. For example, in a manufacturing facility, a robotic arm is responsible for handling hazardous chemicals on a conveyor belt under the guidance of an operator in local space (see Fig.\ref{fig:teaser} lower right of the center schematic). The remote space is characterized by its task-oriented property, utilizing the physical capabilities of the robot to perform specific actions that benefit from or require teleoperation.

To provide a comprehensive understanding of the two space types mentioned above, we have conducted a detailed synthesis and analysis of the relevant literature. Based on this, we have categorized the system design within this domain along several dimensions: 1) the \textbf{XR technologies} approach used in remote HRI; this dimension explores the use of various XR technologies supported in local space -- such as VR, AR, and MR -- to bridge the gap between local and remote spaces, providing virtual manipulation interfaces for local users. Our focus is on how these technologies can be utilized to create immersive and intuitive interfaces to control robots in remote space locations. 2) the \textbf{interaction modalities} between the user and the virtual interface; interaction modalities are methods by which users communicate with and control virtual interfaces. This includes gesture control, controller control, and motion capture, etc.  The choice of interaction method influences the efficiency of the user's manipulation of the virtual interface, which in turn affects the robot's ability to accurately and efficiently perform tasks in the remote space, emphasizing the design requirements of catering to the context of the task and the user's physical environment in the local space. 3) the design of the \textbf{virtual interface}; this dimension examines how information in remote space is presented in local space (e.g., a digital twin of a robot in a remote space or a 3D reconstruction of a remote environment), and the interface design may need to take into account the cognitive load of the user and strike a balance between complexity and usability. 4) the \textbf{user's perspective} of observing the robot's actions in the remote space; This dimension is concerned with how the user perceives and understands the robot's movements and remote environment through the virtual interface. It deals with the perspective from which the user observes the robot or remote space (coupled to the robot, decoupled from the robot, or \textcolor{black}{bird's-eye view}, etc.) and how this perspective enhances or hinders task execution. The choice of perspective affects the user's situational awareness, the design of the virtual interface, and the intuitiveness of controlling the robot. 5) the \textbf{robot and specific tasks classification} in the remote space; this dimension involves categorizing robots based on their physical capabilities and the tasks they perform. Understanding this dimension helps developers customize the virtual interfaces and user perspectives of XR technologies to meet the specific needs of robots and tasks. and 6) the \textbf{enhancement locations and types} of the multimodal elements; this dimension examines how multimodal elements (e.g., visual, auditory, and haptic feedback) can be augmented and integrated into local space and remote space to improve user control and perception of the robot. Enhancement types may include highlighting important control or feedback cues (e.g., text, video, 3D graphic overlays) and providing simulated haptic feedback to emulate the physical interaction of objects in the remote space.

\subsection{Existing Surveys}

Previous research has conducted separate investigations into XR technology, remote collaboration, and HRI. However, the intersection of these three dimensions, particularly remote HRI, has remained largely unexplored. Notably, Schafer et al.~\cite{schafer2021survey}, and Wang et al.~\cite{wang2021ar} have conducted reviews of remote collaboration systems using XR technology. Schafer et al. emphasized synchronous remote collaboration systems, whereas Wang et al. concentrated on physical tasks. However, their main focus is human-to-human remote collaboration instead of human-robot interaction. Moreover, when analyzing HRI systems or collaborative robots, the majority of researchers have primarily explored the application of AR technology~\cite{suzuki2022augmented,bassyouni2021augmented,makhataeva2020augmented,costa2022augmented,de2020systematic,green2007human,phaijit2022taxonomy}. The study by Dianatfar et al.~\cite{dianatfar2021review} encompasses VR technology but only synthesizes VR simulation applications for surgical robots and does not adequately consider interaction scenarios between humans and tangible robots. Walker et al.~\cite{walker2022virtual} proposed a taxonomy for HRI systems using XR technology, but their primary focus was not HRI in remote contexts.

In reviewing the existing literature, we notice that none of the existing survey articles systematically categorize and synthesize the usage of XR technologies in remote HRI settings. In contrast, our article addresses the gaps in the current research, including all XR technologies, and dives into the issue of HRI in remote contexts. This systematic review can serve as the first comprehensive guide for researchers to situate their work within a broader framework and explore innovative systems for XR-based remote HRI.

\subsection{Structure of the Survey}

The remainder of this paper is structured as follows: Section~\ref{sec:methodology} thoroughly explains the methodology employed for this survey. Section~\ref{sec:result} offers an in-depth discussion and analysis of the included articles, specifically focusing on the utilized techniques, types and tasks of remote robots, task evaluation, the role of XR techniques, multiplayer/robot support, and system latency. This analysis is based on our developed taxonomy and data extraction rules. In Section~\ref{sec:discussion}, a detailed discussion and analysis concerning the development and recent advancements in XR-based remote HRI are presented. Subsequently, Section~\ref{sec:future} delves into the challenges to the research and suggests potential future research directions. Finally, Section~\ref{sec:conclusion} provides a comprehensive summary of the entire paper.
\section{Methodology}
\label{sec:methodology}

\begin{figure}[tbh!]
  \centering
  \includegraphics[width=.6\linewidth]{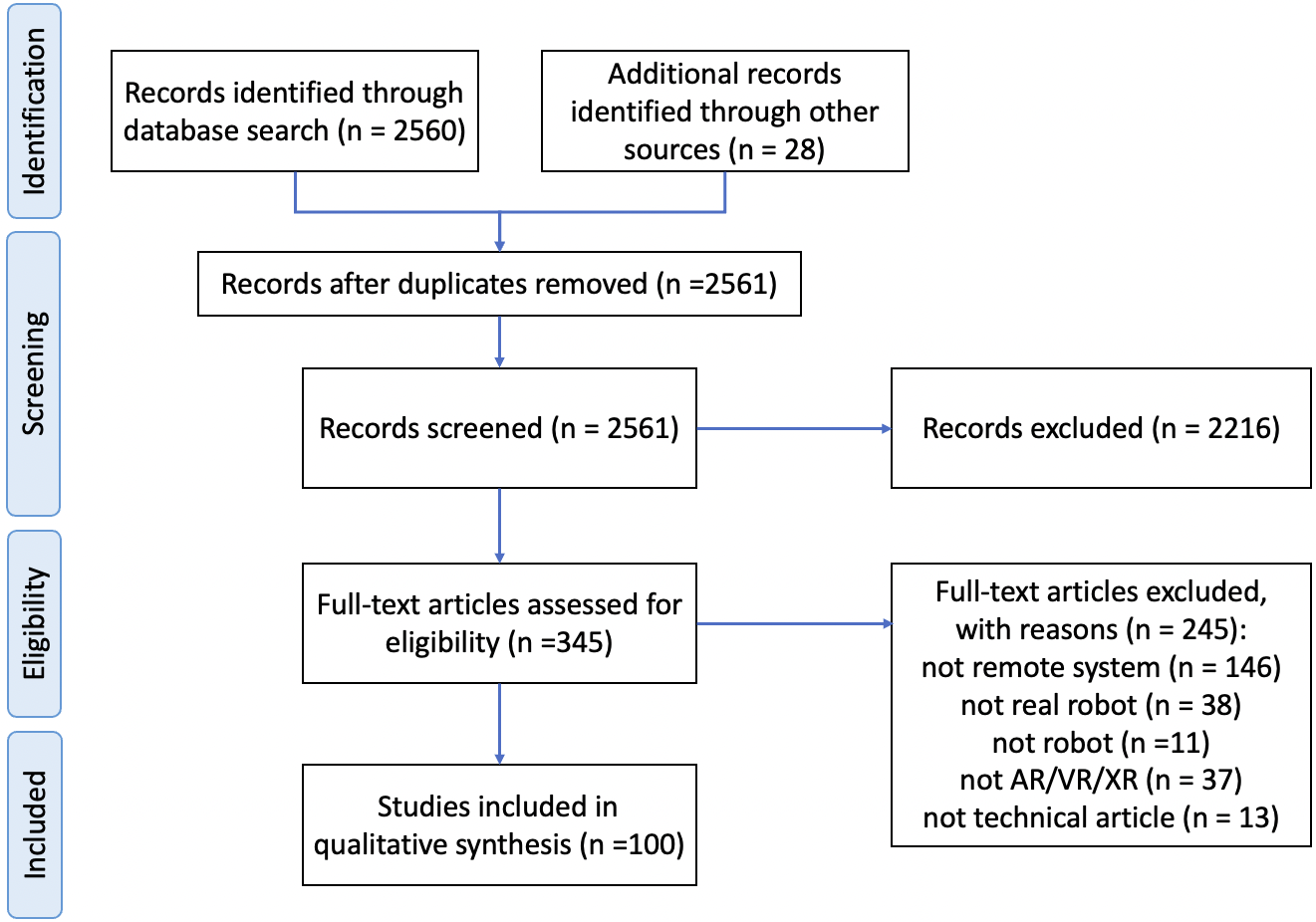}
  \caption{Systematic review process using PRISMA.}
  \label{fig:prisma}
\end{figure}

To ensure methodological rigour and transparency in our literature review process, we utilized the Preferred Reporting Items for Systematic Reviews and Meta-Analyses (PRISMA) framework, as recommended by Takkouche et al.~\cite{takkouche2011prisma}. The review was conducted collaboratively by multiple authors, using an online tool named \textbf{Covidence\footnote{\url{https://www.covidence.org/}}}. The complete PRISMA results can be found in Figure~\ref{fig:prisma}. Upon completing our search, we began a structural process of filtering through 2,588 articles. We initially removed 27 duplicate articles, leaving us with a pool of 2,561 articles to examine. We screened these articles based on their titles and abstracts, leading to the exclusion of 2,216 articles that did not meet our criteria. Following a full-text evaluation, an additional 245 articles were further excluded. The specific inclusion and exclusion criteria are described in Section~\ref{sec:criteria}. 
Our literature review process was conducted in two phases to capture the most current research. The first round of data extraction took place in May 2022, followed by a second round in December 2023. This two-phase approach allowed us to incorporate the latest studies and ensure a comprehensive review of the literature up to December 2023. Ultimately, we selected 100 articles for data extraction and further analysis in our survey.

\subsection{Search Strategy}
\subsubsection{Keywords}

 The keywords primarily reflect the three dimensions of remote operations, human-robot interaction and VAM technology. Considering the terminological expressions in different contexts, we selected the keywords ``distributed'', ``remote'', ``teleoperation'',  ``telerobotics'', ``telepresence'' and ``spatial'' for the remote environment. The main keywords involved in human-robot interaction are ``robotic'', ``robot'', ``machine'', ``human-robot interaction'', ``human-robot collaboration'', ``interaction'', ``collaboration'', ``cooperation'', ``collaborate'' and the abbreviations ``HRI'' and ``HRC'', the keywords and abbreviations related to VAM technology we have chosen ``virtual reality'', ``augmented reality'', ``mixed reality'', ``vr'', ``ar'' and ``mr''.
 
\subsubsection{Databases}

To ensure a comprehensive literature review, we searched for the most relevant articles available on several publication databases, including ACM Digital Library, IEEE Xplore, and ScienceDirect (Elsevier). To supplement this search, we also conducted a snowball search on Google Scholar, which involved reviewing the reference lists of relevant articles to identify additional publications. Additionally, we utilized the \textbf{Connected Papers\footnote{\url{https://www.connectedpapers.com/}}} online tool to identify related articles and broaden our search until no new relevant articles appeared. This approach allowed us to thoroughly explore the relevant literature and identify key publications related to our research question.

\subsection{Inclusion and Exclusion Criteria}
\label{sec:criteria}

During the Screening and Eligibility stages of our PRISMA review, we carefully developed appropriate inclusion and exclusion criteria to guide our selection process. We have summarized these criteria in Table~\ref{tab:inclusion}, which outlines the factors that we used to identify related work of interest. Our analysis focused on articles that met one or more of the inclusion criteria, while also ensuring that any articles that met at least one of the exclusion criteria were removed from consideration. By adhering to these guidelines, we did our utmost to ensure a rigorous and thorough review of the relevant literature.

{
\footnotesize
\begin{table}[tbh!]
\caption{Inclusion and Exclusion Criteria}
\resizebox{1\linewidth}{!}{ 
\begin{tabular}{p{60pt}|p{400pt}}
\toprule
Criterion & Description                                                             \\ \hline
$I_1$ & The research proposes the design, development or system for remote control of robots based on augmented, virtual or mixed reality. \\
$I_2$ & The research used augmented, virtual or mixed reality as a method of remote control of the robot.                                  \\ \hline
$E_1$         & The system proposed in the study does not support remote control.       \\
$E_2$         & The study does not involve real robots.                                 \\
$E_3$        & The target of research control is not a robot.                          \\
$E_4$         & The study did not use augmented, virtual or mixed reality technologies. \\
$E_5$         & This is not a technical article.                                        \\ \bottomrule
\end{tabular}}
\label{tab:inclusion}
\end{table}
}

\subsection{Data Extraction and Analysis}

To extract relevant information from the included articles, we developed a data extraction rubric that enabled us to systematically capture the key aspects of augmented, virtual or mixed reality technologies, Human-robot interaction, and remote control. Initially, the first author of this survey selected 10 articles in a pseudo-random manner (\cite{walker_robot_2019,jang_omnipotent_2021,grzeskowiak_toward_2020,bian_interface_2018,wang_human-centered_2019,kuo_development_2021,bai_kinect-based_2018,xie_framework_2022,trejo_user_2018,zhao_robot_2017}), and developed items based on the relevant aspects identified in the articles. The initial data extraction rubric was then evaluated by all authors and refined into the final version described in Table~\ref{tab:dataExtraction}. The first and second authors independently extracted data from each article using this finalised rubric. In cases of conflicting data, consensus was reached through discussions among the authors.

Data extraction items \textbf{DE1-DE3} pertain to general descriptors of the paper, including study number (author and date), title, and keywords. XR technologies (\textbf{DE4}) encompass common types of augmented, virtual, and mixed reality technologies, such as VR HMD, VR video, AR HMD, Mobile AR, projectors, MR, CAVE, and others. Interaction modalities (\textbf{DE5}) highlight the variety of hardware and their applications in virtual environments. The telepresence interface (\textbf{DE6}) allows users to operate the robot more intuitively with a virtual interface. It may include a direct interface, a digital twin of the remote robot, a virtual control room, or a digital twin of the remote robot combined with a 3D reconstruction of the remote environment. The user's perspective (\textbf{DE7}) refers to the user's viewpoint, such as being bound to the robot's perspective, observing the robot from a detached third view or a top-down "God's view," or having the ability to change perspectives based on need. Generic types of robots are described under \textbf{DE8}, while specific tasks (\textbf{DE9}) that can be performed by the human-robot collaboration system are of interest, as different tasks might require various robots or telepresence designs. Realizing the full potential of the XR also intends to improve the efficiency and intuitiveness of remote control. The system may have been enhanced in different locations (\textbf{DE13}) with multimodal enhancements (\textbf{DE14}), e.g., haptic, video, 2D or 3D overlays, avatars, and more. In addition, human-robot interaction may include multiplayer or robot collaboration (\textbf{DE12}), necessitating distinctions between one-to-one, one-to-multi, or multi-to-multi collaboration, which could influence study design. The evaluation method of the study (\textbf{DE10}) is also essential; it could be quantitative, qualitative (e.g., questionnaires), or focused on the robot rather than the user, including the potential presence of delays (\textbf{DE11}) in remote controls.

{
\footnotesize
\begin{table}[tbh!]
\caption{Data Extraction Rubric for the Selected 100 Articles}
\resizebox{1\linewidth}{!}{ 
\begin{tabular}{p{15pt}p{120pt}p{300pt}}
\toprule
\rowcolor[rgb]{0.851,0.851,0.851} ID & Data Extraction                                        & Type\\                                                                                                                                               
\midrule
DE1                                  & Study ID                                               & Open text                             \\
DE2                                  & Title                                                  & Open text                              \\
DE3                                  & Keywords                                               & Open text    \\
DE4                                  & Used XR technologies                                   & VR HMD,~AR HMD,~MR,~Mobile AR,~CAVE, Other\\
DE5                                    & Interaction modalities                               & Gesture, Controller, Joystick, Gaze, Head, Haptic devices, Motion capture, Walk, 2D screen, Voice, Glove, Other \\
DE6                                 & Virtual interfaces                        & Direct, Digital twin, Virtual control room, Digital twin+3D reconstruction, Direct+3D reconstruction, 3D reconstruction, Context-aware AR, Multiple, Other          \\
DE7                                  & User perspective                              & Coupling with robots, Decoupling from robots, Dynamic perspective, Bird's-eye view, Other  \\
DE8                                  & Type of robots                                         & Mobile robot, Drones/UAV, Humanoid Robot, Robotic arm, Mobile robot+robotic arm, Double-armed robot, Medical Robotics, Other          \\
DE9                                 & Specific tasks involved                                & Navigation, Grabbing/Picking/Placement, Surgery, Game/Entertainment, Industrial/Manufacturing, Search, Environment scan, No, Multiple, Other    \\
DE10                                 & How was it measured or evaluated?                      & Time/accuracy of the task, Interviews, Questionnaire, AR/VR performance, Comparison, N/A, Other\\
DE11                                 & Was there a discussion about delays?                   & No, Yes(Times), Other  \\
DE12                                  & Support multiplayer collaboration?                     & No, Multi-user - one robot, One user - multi-robot, Multi - Multi, Other                                \\
DE13                                & Where are the enhancements located?  & User, Robot (real), Robot (virtual), Object (virtual), Real environment(RE), Virtual environment(VE)          \\            
DE14                                 & Enhancement types & Haptic, Voice, Graphic, Text, 3D Object, Highlight, Ray, Avatar \\
\bottomrule
\end{tabular}}
\label{tab:dataExtraction}
\end{table}
}

\section{Results and Descriptive Statistics}
\label{sec:result}
\subsection{Overview of Included Articles}

\begin{figure}[tbh!]
  \centering
  \includegraphics[width=\linewidth]{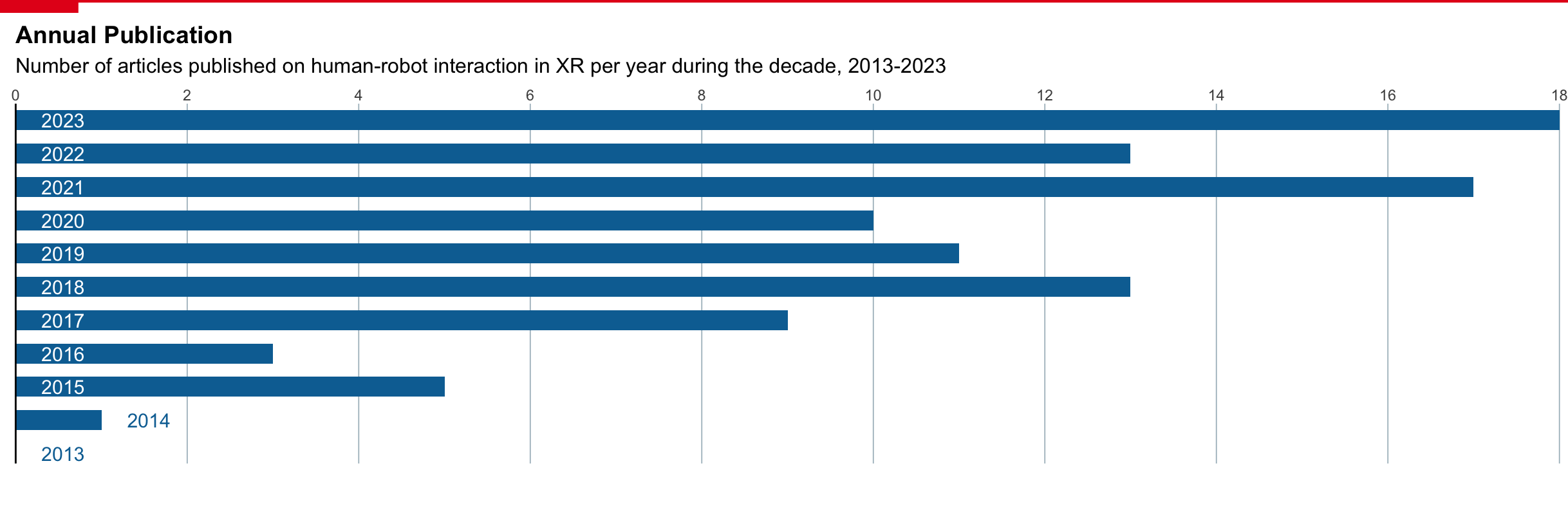}
  \caption{Articles per year (N=100)}
  \label{fig:year}
\end{figure}

We meticulously extracted pertinent information from the 100 articles identified during our screening. The selected articles span from 2013 to 2023, and the annual publication count is illustrated in Figure~\ref{fig:year}. These articles are from well-known venues for HRI and human-computer interaction (HCI), including the IEEE International Conference on Intelligent Robots and Systems (IROS), ACM/IEEE International Conference on Human-Robot Interaction (HRI), IEEE International Symposium on Robot and Human Interactive Communication (RO-MAN), and the ACM Symposium on User Interface Software and Technology (UIST). These data were subsequently analyzed and summarized both statistically and graphically, with additional qualitative insights emerging during the iterative analysis. A comprehensive list of data extracts for the included articles can be found in Appendix~\ref{appendix}.  Over the past decade, there has been a general increase in the number of articles published on the investigated topics, indicating growing interest and significance in recent years. This trend could be attributed to the landscape of XR technology maturing, and it is worth noting that the volume of publications in this field reached its zenith in 2023. 


\begin{figure}[htb]
  \centering
  \includegraphics[width=.4\linewidth]{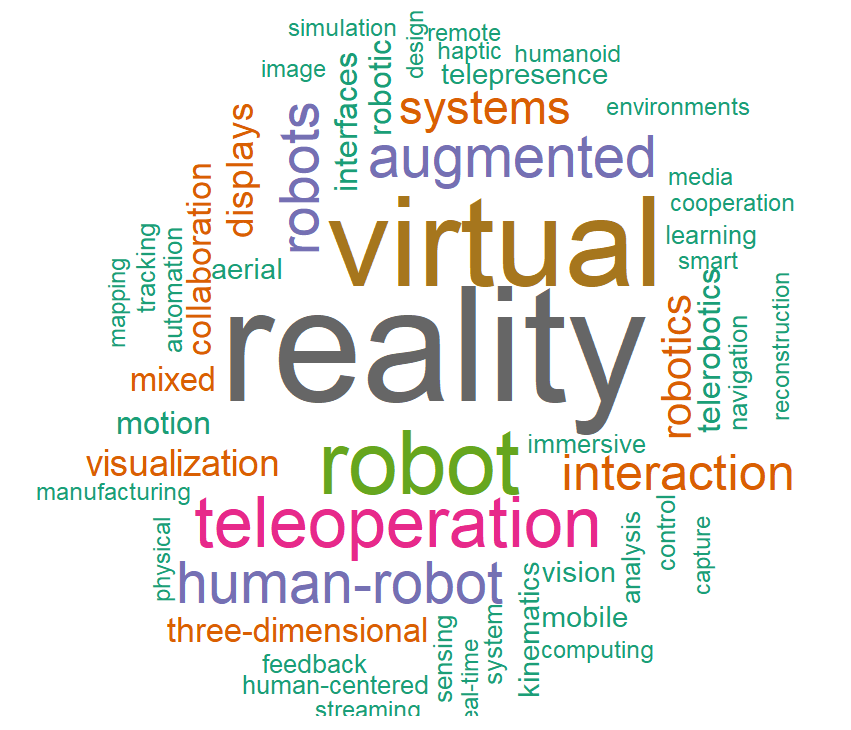}
  \caption{Frequently-used keywords in the included articles}
  \label{fig:wordcloud}
\end{figure}

The articles originate primarily from the robotics and manufacturing domains, and our uniqueness is to extend the domains to remote interaction. Keywords such as ``reality'' and ``virtual'' frequently appear, with ``augmented'' being a high frequency term, although less so than ``virtual''. This observation may suggest a preference for virtual reality over augmented reality in this research area, but a more detailed analysis is needed to confirm this. Keywords such as ``robot'', ``robots'', and ``robotics'' are also prevalent in statistics. The high frequency of the term ``teleoperation'' indicates that these articles predominantly focus on teleoperation, while the keyword ``human-robot'' often appears in the context of collaboration between humans and robots. These two keywords co-exist in several articles, suggesting that they explore the intersection of these two themes, i.e., teleoperated control in human-robot collaboration. Some articles refer to this cooperation as ``collaboration'', while others use the term ``interaction''. Figure~\ref{fig:wordcloud} presents the top 100 keywords in a word cloud, providing a more precise visualization of the frequency distribution of these terms within the included articles.

\subsection{Technologies}

\begin{figure}[htb]
  \centering
  \includegraphics[width=1\linewidth]{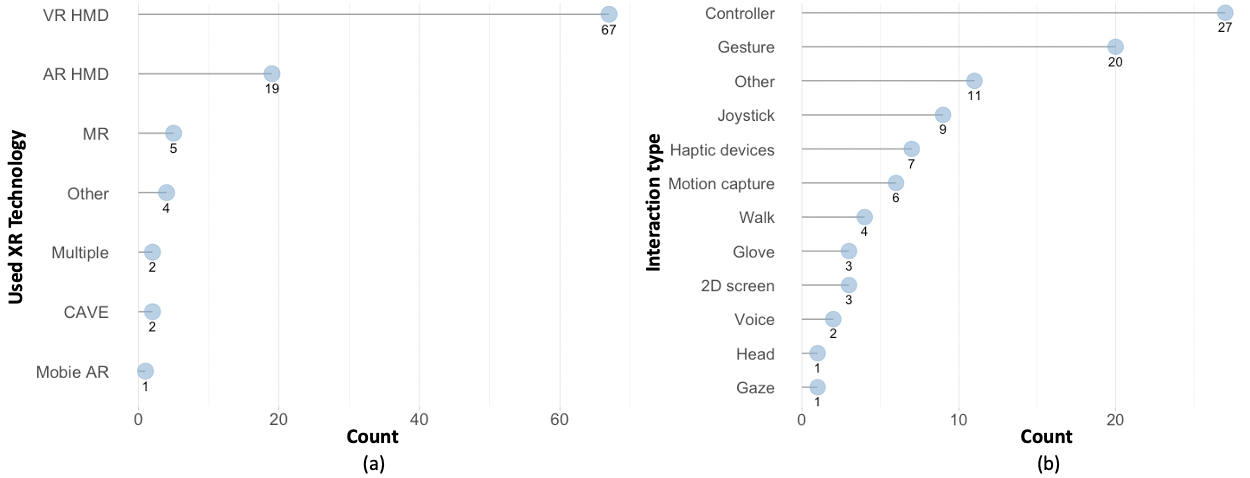}
  \caption{Use of different XR devices and interaction types in the included articles. (a) Statistics on the type of XR technology used in the studies; (b) Statistics on the interaction modality used in the studies.}
  \label{fig:tech}
\end{figure}

Figure~\ref{fig:tech} depicts the employment of assorted XR technologies and diverse interaction modalities in the analyzed studies. It is evident that Virtual Reality Head-Mounted Display (VR HMD) dominates the landscape, constituting 67\% of adoption, and significantly surpassing other apparatuses. Augmented Reality Head-Mounted Displays (AR HMDs) follow with 19\% prevalence, while a limited number of studies utilize MR (5\%), CAVE (2\%), and Mobile AR (1\%) technologies. 

A majority of studies favor VR technology, which can likely be attributed to the necessity for users operating remotely to receive information about the remote robotic system or environment. VR technology delivers this data in a more immersive, intuitive manner, fostering a heightened sense of presence -- a key advantage of VR technology. In contrast, AR technology excels at superimposing virtual information onto real environments. Nonetheless, in the context of remote operation, AR faces challenges in satisfying the demand for information overlay on the real robot and its workspace (where the user is absent), possibly accounting for the higher prevalence of VR technology in research. Another contributing factor could be the lower cost of commercial VR HMDs compared to AR HMDs~\cite{xu2021hmd}. 
Furthermore, VR HMDs can integrate supplementary depth cameras to achieve functionality similar to AR HMDs, as exemplified by the study conducted by Yew et al.~\cite{yew_immersive_2017}. Their research prototype used the attached camera to track the pose of the Oculus Rift HMD and the robotic arm to generate and display the AR environment in the HMD. This factor may explain the increased usage of VR HMDs in research. Additionally, a handful of studies employ a combination of devices, such as AR and VR~\cite{aschenbrenner_collaborative_2018}, often within the context of multi-person remote collaboration. In these scenarios, a local operator utilizes an AR HMD to manipulate the robot, while a remote expert wears a VR HMD to obtain immersive three-dimensional guidance of the local robot and work environment. Subsequently, this guidance information is transmitted to the local worker's AR HMD.

The most prevalent interaction type involves using built-in controllers provided with the devices (27\%). This trend is reasonable, as the most commonly utilized XR devices are VR HMDs and commercial VR HMDs generally include their proprietary controllers, regardless of form factors. Gestural interaction is another frequent method (20\%), primarily because commercial AR HMDs predominantly use hand or head gestures for interaction with virtual content. Additionally, many studies employ VR HMDs with supplementary depth cameras, such as Leap Motion~\footnote{\url{https://www.ultraleap.com/product/leap-motion-controller/}}, mounted on the headset to detect user gestures, as remote robot operation through gestures is often more intuitive than using controllers. The joystick interaction, typically associated with gamepads like Xbox~\footnote{\url{https://www.xbox.com/en-SG/accessories/controllers/xbox-wireless-controller}}, features prominently in the research (9\%) and is often considered when the subject robots are drones or mobile robots~\cite{zein_deep_2021,hedayati_improving_2018,ai_real-time_2016,betancourt_exocentric_2022,walker_robot_2019,wibowo_improving_2021,stotko_vr_2019}, only the study by Vu et al.~\cite{vu_investigation_2022} used the joystick to manipulate the robotic arm. The use of virtual fixture haptic devices is also relatively high (7\%). Haptic devices can overlay enhanced sensory information on users' perception of the real environment to improve human performance in both direct and teleoperated tasks~\cite{fani2018simplifying}. Motion capture interactions often necessitate users wearing sensors (6\%) and mapping robotic arms to arm or shoulder coordinates, rendering operations more intuitive and reducing learning costs. The remaining interaction types -- such as actual walking in remote environments (4\%), 2D screen (3\%), voice (2\%), glove (3\%), head movement (1\%), and gaze (1\%) -- constitute a minor percentage overall. These less common interaction options are frequently linked with specific robot operation tasks. For example, Moniri et al. studied user visual attention in HRI, and gaze was chosen as the interaction method since the human eye gaze is an important indicator of the direction of visual attention focus~\cite{moniri_human_2016}.

\begin{figure}[htb]
  \centering
  \includegraphics[width=1\linewidth]{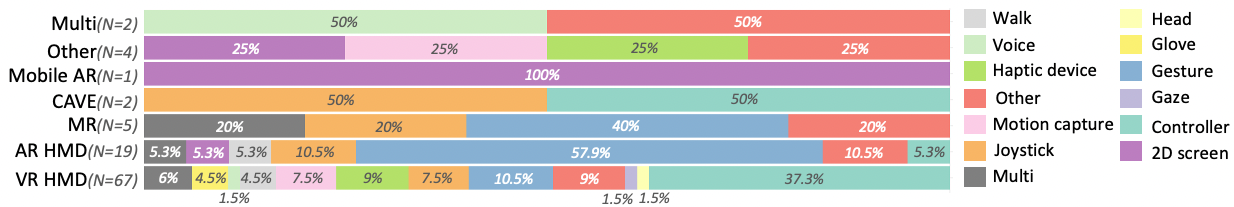}
  \caption{Percentage bar stacking chart for different XR technologies corresponding to the interaction types.}
  \label{fig:devices}
\end{figure}

Building on the previously discussed XR techniques and interaction modalities, Figure~\ref{fig:devices} provides a comprehensive overview of these elements in the examined research. Interaction refers to how users manipulate virtual environments or objects through their actions. Our findings indicate that the choice of XR technology partially determines the interaction paradigm. VR emerges as the most popular option for remotely controlling robots, with most studies opting for controllers or gestures as interaction methods. VR devices appear compatible with a diverse range of control techniques, except for 2D screens (i.e., touchscreen devices such as tablets or smartphones), typically not employed by VR HMDs. This exclusion is logical, given that VR devices obstruct the user's line of sight to the real world, making it impossible for users to view content on a 2D screen while wearing a VR HMD~\cite{burdea2003virtual}. In contrast, most systems utilizing AR HMDs rely on gestures for interaction~\cite{funk2017hololens}. This preference may stem from the nature of consumer-grade AR HMDs, in general, not including proprietary controllers, making gestures a convenient, self-contained interaction solution.

\subsection{Robot Types and Tasks}
\label{sec:robotTypesTasks}

\begin{figure}[htb]
  \centering
  \includegraphics[width=1\linewidth]{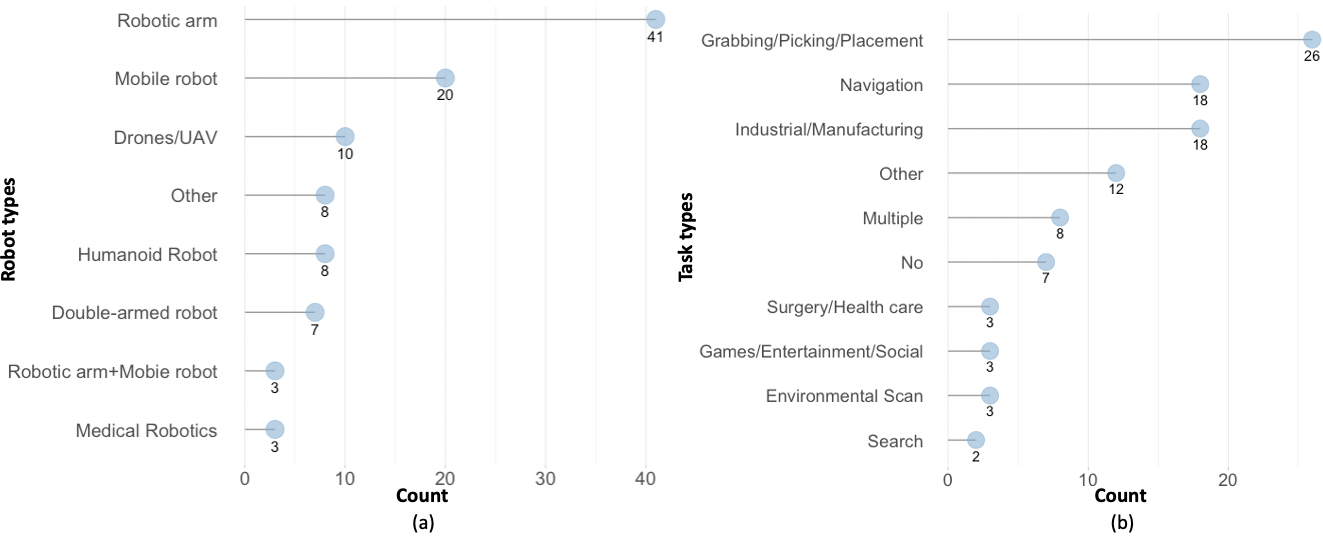}
  \caption{Different robots and different types of tasks included in the study. (a) Statistics of the various robot types used in the reviewed studies; (b) Statistics on tasks performed by robots.}
  \label{fig:robots}
\end{figure}

\begin{figure}[htb]
  \centering
  \includegraphics[width=1\linewidth]{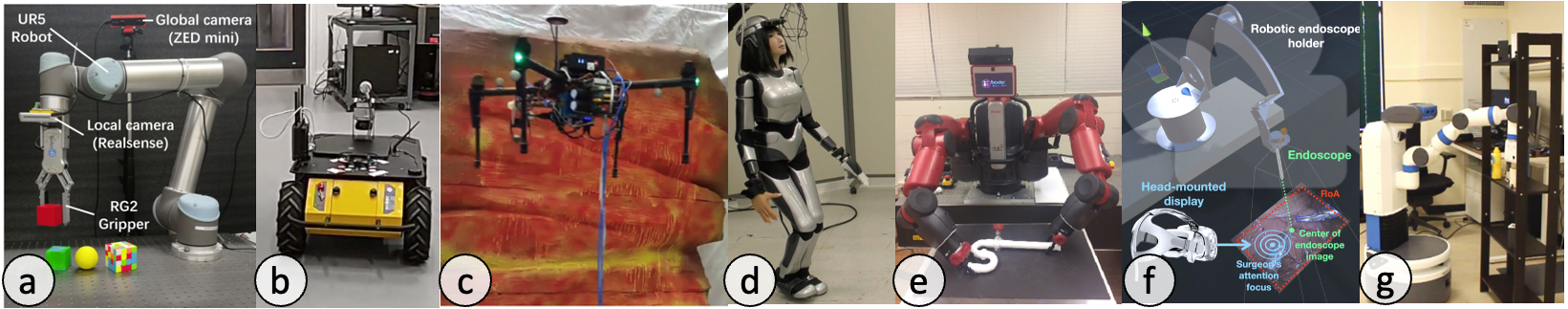}
  \caption{Examples of different robot types: a) Robotic arm~\cite{wei_multi-view_2021}; b) Mobile robot~\cite{stedman_vrtab-map_2022}; c) Drones/UAV~\cite{vempati_virtual_2019}; d) Humaniod robot\cite{chen_enhanced_2022}; e) Double-armed robot~\cite{zhou_intuitive_2020}; f) Medical robot\cite{zinchenko_autonomous_2021}; f) Robotic arm + mobile robot~\cite{hernandez_increasing_2020}.}
  \label{fig:robotsExamples}
\end{figure}

Our analysis offers a summary of the various robot types (see Fig.~\ref{fig:robotsExamples} with examples) featured in the included studies, as well as the specific tasks discussed or tested in the papers. This information is illustrated in Figure~\ref{fig:robots}. As shown in Figure~\ref{fig:robots}(a), the robotic arm is the type of robot most extensively researched, accounting for 41\% of the studies. Mobile robots and drones follow, with respective shares of 20\% and 10\%. We categorize a special type of robot -- robotic arm + mobile robot. This type of robot has the characteristics of both robotic arms and mobile robots, and its main structure is a movable base on which there is a robotic arm. It can be interpreted as a movable robotic arm. Such a robot type was 3\% of the included articles. Other types of robots, such as humanoid robots (with human facial features, 8\%), two-armed robots (7\%), and medical robots (3\%), constitute a smaller portion of the overall robot types. Regarding the specific tasks performed by the robots, Figure~\ref{fig:robots}(b) reveals that the three most dominant task types are object grasping/picking/placement, robot navigation, and specialized operations in industry/manufacturing, representing 26\%, 18\%, and 18\% of the task types, respectively. A notable fraction of task types is multiple (8\%), while some studies do not specify the exact tasks that the robots in the research can execute (7\%). The remaining types of tasks, such as remote environmental scanning (3\%), surgery/ health care (3\%), search (2\%), and gaming/ entertainment (3\%), comprise a minimal percentage.

\begin{figure}[htb]
  \centering
  \includegraphics[width=1\linewidth]{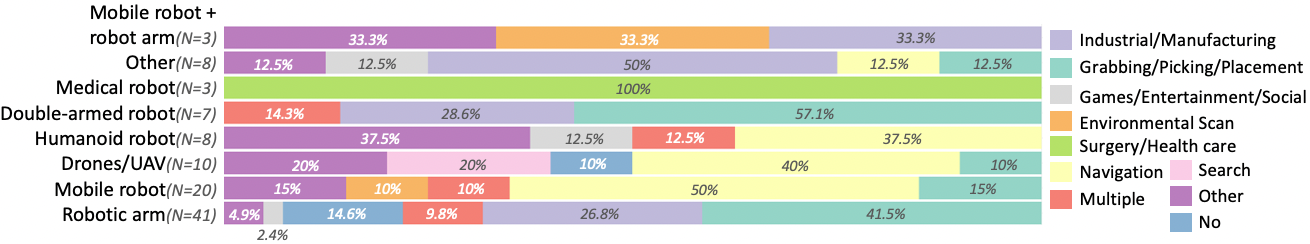}
  \caption{Percentage stacked bar charts for different robots and corresponding task types.}
  \label{fig:tasks}
\end{figure}

We also observed potential correlations between various robot types and task types, as depicted in Figure~\ref{fig:tasks}. Our analysis indicates that robotic arms, two-armed robots, and other specific robot types, such as industrial machines~\cite{mourtzis_augmented_2017,mourtzis_augmented_2019}, maintenance robots~\cite{yew_immersive_2017}, and mining robots~\cite{xie_framework_2022}, are predominantly employed in industrial or production tasks. These robots typically lack mobility capabilities (i.e., they do not possess a chassis that enables movement), so their task types do not involve navigation. Conversely, mobile robots, drones, and a subset of humanoid and other more mobile robots are responsible for tasks such as navigation, search, and environmental scanning that rely on mobility. Robotic arms and two-armed robots primarily perform the grasping, picking, and placement tasks, as these functions connect to the services or fundamental functions of robotic arms. The tasks are also a key component of the industrial or production chain for which robotic arms are originally responsible. To some extent, two-armed robots can be considered a combination of two robotic arms. Medical robots represent the most homogeneous robot type, as their sole responsibility is to assist in surgical procedures~\cite{zinchenko_autonomous_2021,trejo_user_2018}. Furthermore, we discovered that humanoid robots appear to possess unique social characteristics. The tasks they are assigned, such as intervening with children diagnosed with autism spectrum disorders (ASD)~\cite{kulikovskiy_can_2021}, engaging in chess games with remote users~\cite{schwarz_nimbro_2021}, assisting remote users with dressing~\cite{schwarz_nimbro_2021}, expressing emotions~\cite{takacs_towards_2015}, receiving and guiding users~\cite{grzeskowiak_toward_2020}, and maintaining road traffic security~\cite{gong_real-time_2017}, are inherently linked to human or social activities. This suggests that humanoid robots, due to their anthropomorphic form and capabilities, are particularly suited for roles that require social interaction or human-like tasks.

\subsection{Evaluation of Tasks}
\label{sec:EvaluationOfTask}
\begin{figure}[htb]
  \centering
  \includegraphics[width=.5\linewidth]{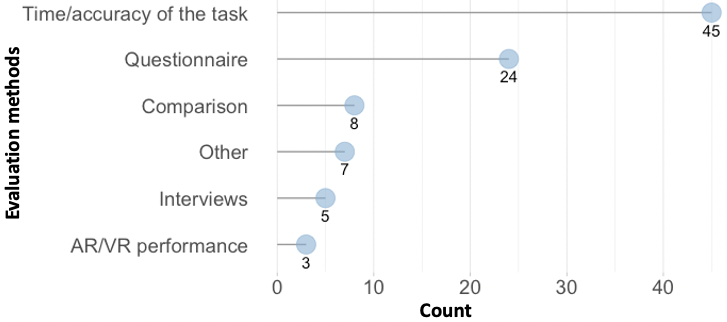}
  \caption{Statistics of different evaluation methods.}
  \label{fig:evaluation}
\end{figure}

We examined the evaluation methods used in the included studies. We discovered that 38\% of the papers did not conduct any form of evaluation, while 36\% employed a single evaluation method, and 36\% utilized a hybrid approach. Among these methods, 69\% of the papers employed a quantitative evaluation, which included assessing time/accuracy to complete the task (48.91\%), administering standardized questionnaires such as the NASA-TLX~\cite{hart1988development}, or employing task-based design questionnaires (26.09\%). A mere 5.43\% of the papers focused on qualitative user evaluations, such as interviews. Additionally, a small subset of papers, amounting to 8.70\%, concentrated on comparing the performance of robots and their digital twins. This type of study is typically evaluated by comparing the trajectory coordinates of the system's input and the robot's output. For instance, studies by Yun et al.~\cite{yun_immersive_2022}, Cousins et al.~\cite{cousins_development_2017}, and Bian et al.~\cite{bian_interface_2018} evaluated the coordinates of the user's hand input and the robot arm's output. Betancourt et al.'s study compared the 3D spatial coordinates of a virtual drone and a real flying vehicle~\cite{betancourt_exocentric_2022}. A few studies, constituting 3.26\%, evaluated the performance or impact of AR/VR itself. For example, Kuo et al. compared the accuracy of manipulating objects through VR, video, and the real world~\cite{kuo_development_2021}. Similarly, Chen et al. evaluated different methods of 3D reconstruction in VR~\cite{chen_real-time_2020}.  

\subsection{XR Technologies Facilitate Effective Remote Robot Collaboration}
\label{sec:InterfaceEnhancedResult}
\subsubsection{Virtual Interface Design}
\label{sec:VirtualInterfaceResult}

\begin{figure}[htb]
  \centering
  \includegraphics[width=1\linewidth]{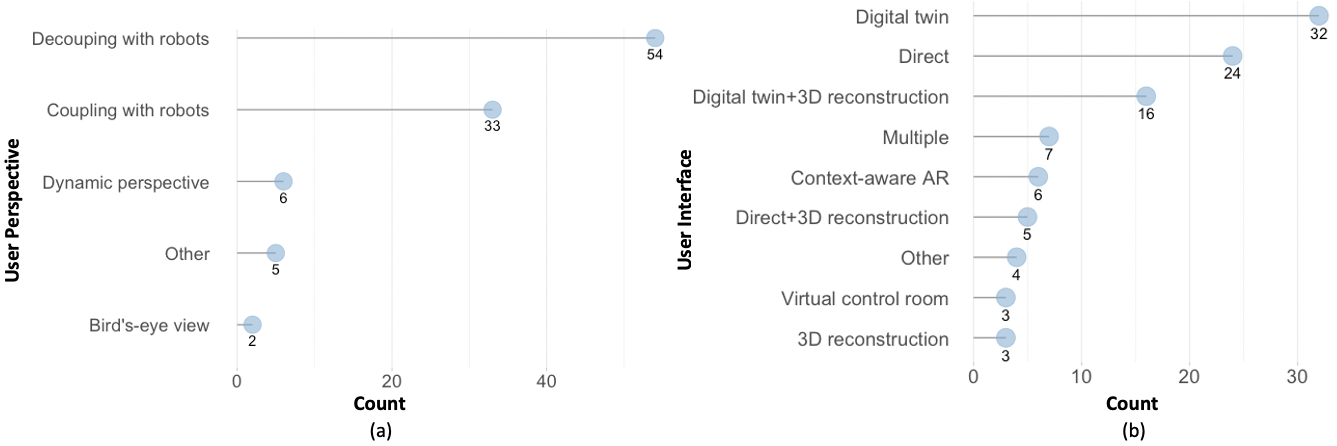}
  \caption{(a) Categorization and statistics of user perspectives; (b) Statistics on the number of different types of user interfaces in the review.}
  \label{fig:pov}
\end{figure}

We adopt the taxonomy proposed by Walker et al.~\cite{walker2022virtual} to analyze the user interface design in XR, dividing it into two components: user perspective and user interface (see Figure~\ref{fig:pov}). The user perspective comprises five categories: robot-coupled, robot-decoupled, Bird's-eye view, dynamic perspective, and others. Robot-coupled perspectives involve users viewing the scene through the ``eyes'' of the robot. This approach was exemplified in the works of Vempati et al.~\cite{vempati_virtual_2019}, Chacko et al.~\cite{chacko_augmented_2020}, and Brizzi et al.~\cite{brizzi_effects_2018}, who conducted their studies in the remote operating systems of UAVs, humanoid robots, and double-armed robots, respectively, with perspectives coupled to the robots.  In this approach, the user's viewpoint is linked to the robot and changes as the robot moves. Conversely, robot-decoupled perspectives enable users to observe the robot's actions from a detached viewpoint, unbound from the robot's movements. This perspective was demonstrated in the work of Kuo et al.~\cite{kuo_development_2021} and Zinchenko et al.~\cite{zinchenko_autonomous_2021}, who developed systems that manipulate a remote robotic arm in VR with a perspective decoupled from the robot. Similarly, Stedman et al.~\cite{stedman_vrtab-map_2022} employed a decoupled perspective in their work with a remote mobile robot. The Bird's-eye view offers an overhead, top-down view, as illustrated by Jang et al.~\cite{jang_virtual_2021}, who utilized this perspective to control swarm robots. while the dynamic perspective allows users to switch between different viewpoints, this perspective was exemplified in the works of Wei et al.~\cite{wei_multi-view_2021} and Xu et al.~\cite{xu_design_2022}, who both employed a combination of two perspectives in their studies. 

\begin{figure}[htb]
  \centering
  \includegraphics[width=1\linewidth]{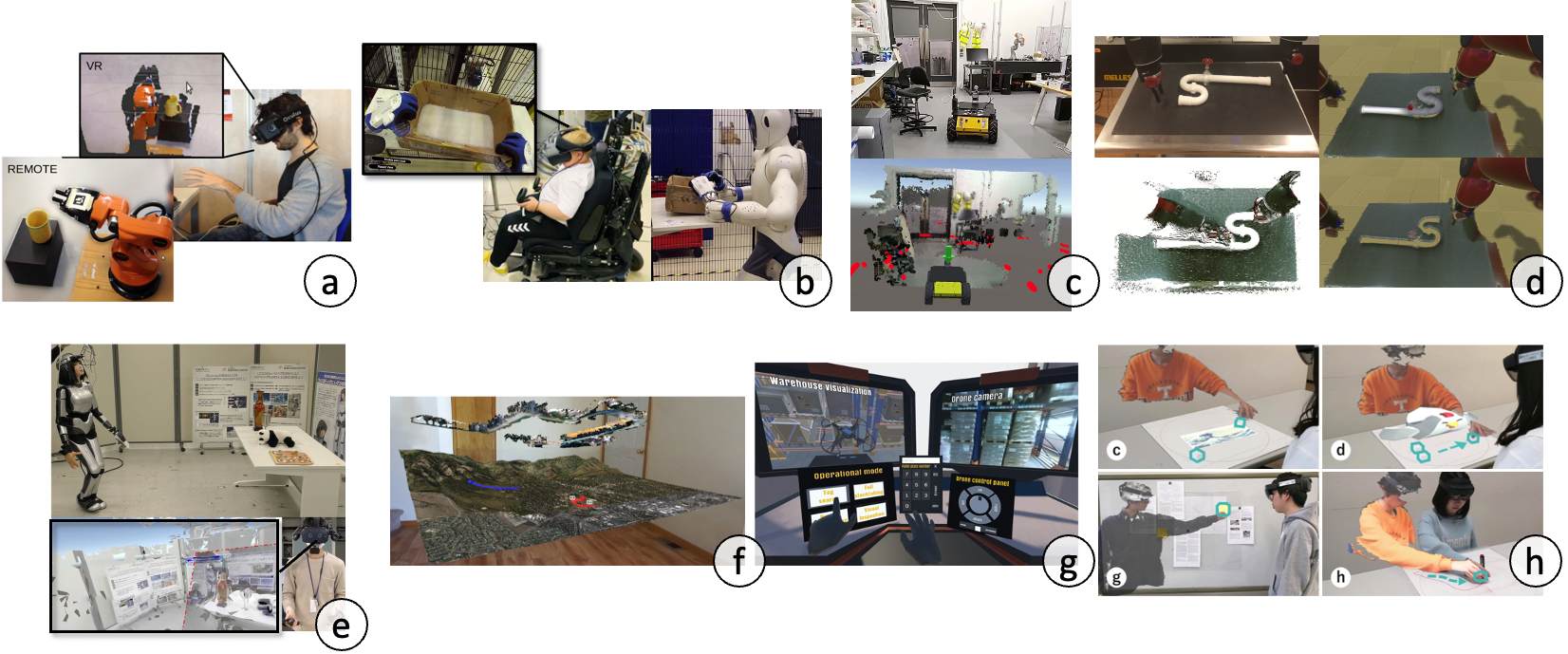}
  \caption{Examples of different virtual user interfaces: a) Digital twin\cite{peppoloni_immersive_2015}; b) Direct interface\cite{sanfilippo2022mixed}; c) Digital twin + 3D reconstruction\cite{stedman_vrtab-map_2022}; d) Multiple interface\cite{zhou_intuitive_2020}; e) Direct + 3D reconstruction interface\cite{chen_enhanced_2022}; f) 3D reconstruction interface\cite{walker_mixed_2021}; g) Virtual control room\cite{kalinov_warevr_2021}; h) Context-aware AR interface\cite{ihara2023holobots}.}
  \label{fig:interfaceExamples}
\end{figure}

We categorize the user interface into several types (see examples in Fig.~\ref{fig:interfaceExamples}). One of these is the direct interface, where the camera on the remote robot side transmits to a 360-degree video interface in the virtual environment. This type of interface, as exemplified by the work of Zhao et al.~\cite{zhao_robot_2017}, allows users to directly observe the remote workspace. Another variant of the direct interface is augmented by a 3D reconstruction of the remote environment, as demonstrated in the work of Chen et al.~\cite{chen_enhanced_2022}. This approach enhances the user's perception of the remote workspace by providing a more immersive and spatially accurate representation. A different approach involves the use of a digital twin of the robot. In this setup, a digital replica of the remote robot exists in the virtual environment, and the user controls the remote robot by manipulating the digital twin, as illustrated in the work of Zinchenko et al.~\cite{zinchenko_autonomous_2021}. Other interface options include the digital twin combined with a 3D reconstruction of the remote environment as an example by Kuo et al.~\cite{kuo_development_2021}, virtual control room -- rebuilds a virtual console in the virtual environment, as demonstrated by Kalinov et al.~\cite{kalinov_warevr_2021}, 3D reconstruction of the remote environment as shown in the work of Zein et al.~\cite{zein_deep_2021}, or a combination of the aforementioned interfaces. These various interface designs offer different levels of immersion, control precision, and spatial awareness, catering to different user needs and task requirements.


\begin{figure}[htb]
  \centering
  \includegraphics[width=1\linewidth]{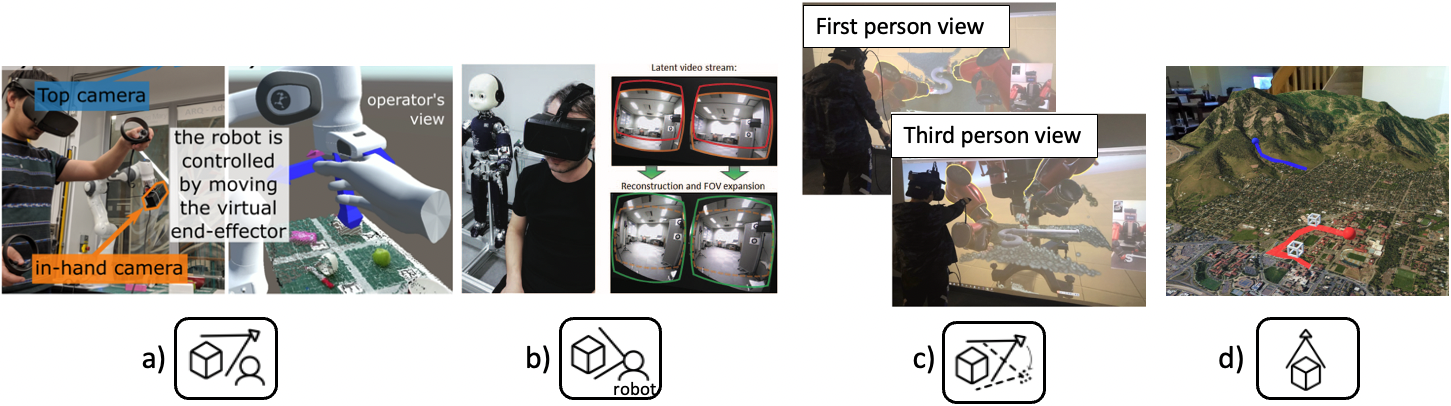}
  \caption{Examples of different user perspectives: a) Decoupling with robot~\cite{omarali_virtual_2020}; b) Coupling with robot~\cite{theofilis_panoramic_2016}; c) Dynamic perspective~\cite{zhou_intuitive_2020}; d) Bird's-eye view~\cite{walker_mixed_2021}.}
  \label{fig:povExamples}
\end{figure}

 We discovered that the user's perspective (see Fig.~\ref{fig:povExamples} with examples) predominantly involves robot-coupled (33\%) or robot-decoupled (54\%) views, while other perspectives, such as dynamic (6\%), bird's-eye view (2\%), and others (5\%), constitute a relatively small proportion. Regarding user interfaces, the digital twin (32\%) and direct interfaces (24\%) are most prevalent, followed by a considerable share of digital twin combined with 3D reconstruction of the remote environment (16\%). The remaining interfaces, including combinations of multiple interfaces (7\%), overlaying virtual interfaces on real environments (i.e., context-aware AR interface, 6\%), direct interfaces augmented with 3D reconstructions of the remote environment (5\%), standalone 3D reconstruction of the remote environment (3\%), virtual control rooms (3\%), and other interfaces (4\%), represent a relatively minor portion.

\begin{figure}[htb]
  \centering
  \includegraphics[width=1\linewidth]{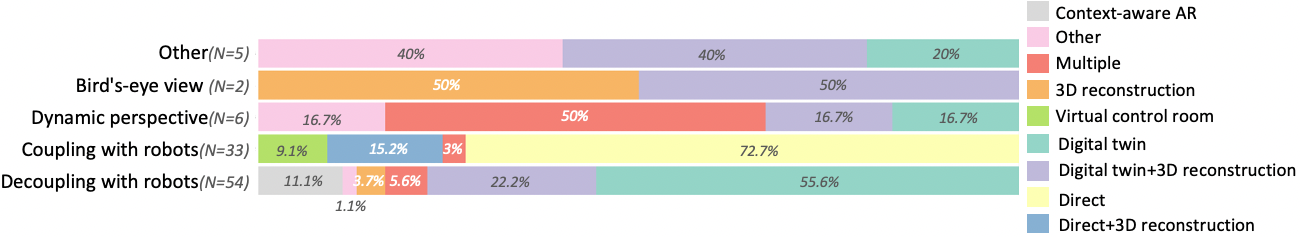}
  \caption{Percentage stacked bar chart of the relationship between the user perspective and the user interface.}
  \label{fig:interface}
\end{figure}

Consistent with our previous analysis, we investigated the relationship between user perspective and user interface using a percentage stacked bar chart, as illustrated in Figure~\ref{fig:interface}. The most significant observation is that the direct user view within the user interface tends to be robot-coupled. This is primarily because users need to observe the remote environment's workspace directly from the robot's viewpoint. Moreover, we found that when the user perspective is decoupled from the robot, it is frequently necessary to incorporate a digital twin of the robot within the XR environment. This is understandable since users need to be aware of the remote robot's motion state, necessitating the creation of a corresponding digital twin in XR to facilitate better comprehension of the robot's operational state. Notably, most studies have opted for the digital twin solution, while only a few, such as Xu et al.~\cite{xu_novel_2018}, have employed a camera positioned next to the remote robot to convey the robot's work status via live video. The virtual control room user interface also requires the user's perspective to be robot-coupled. Although only two studies employed Bird's-eye perspectives, we observed that both execute 3D reconstructions of the remote environment. Lastly, we also discovered that user interface designs for dynamic perspectives tend to be more intricate, such as Zhou et al.~\cite{zhou_intuitive_2020}, often incorporating multiple user interfaces.

\subsubsection{Enhancement}
\label{sec:EhancementResult}
\begin{figure}[htb]
  \centering
  \includegraphics[width=1\linewidth]{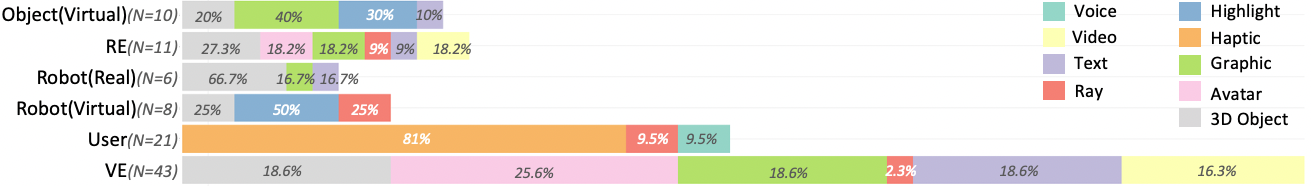}
  \caption{Relationships between types and locations of multimodal enhancements.}
  \label{fig:enhancement}
\end{figure}

In our analysis of the included studies, we evaluated the location and type of information enhanced by the multimodality of the systems. We found that 39\% of the studies employ only a single modal approach to improve remote operations. Most of these systems only supported users to view remote robots, workspaces, or environments immersively via XR. Such systems do not have multimodal enhancements in any location, and the improvement to the user is simply the immersion of XR. On the other hand, 62\% of the studies opted for multimodal enhancement during specific parts of remote robot operation. The primary areas of enhancement were the virtual environment (43.43\%) and the user (21.21\%). A smaller number of studies chose to enhance virtual objects manipulated within the virtual workspace (10.10\%), virtual (8.08\%) and real robots (6.06\%) or in real environments (11.11\%). This could be attributed to the fact that fewer remote operating systems are using AR technology compared to VR technology. AR technology is the predominant technology applicable to augmentation in real robots and real environments. Augmentation on top of real robots often necessitates collaboration among multiple individuals. For instance, the work of Mourtzis et al.~\cite{mourtzis_augmented_2019} and Schwarz et al.~\cite{schwarz_nimbro_2021} involved a cooperative system with multiple users, with one user at a remote location and another user with the robot. However, this type of system represents a very small percentage of our collected papers (for more details, see Section~\ref{sec:multi}). Additionally, we observed that the highlight enhancement mode was only applied to the 
virtual representations of the robot and the objects being manipulated. 
This enhancement type accentuates specific parts of the digital twin or the virtual object for clearer interaction. Avatar enhancement also appeared exclusively in the virtual environment. On the user side, enhancement primarily involved haptic feedback, utilizing tools such as haptic gloves or virtual fixtures. The research conducted by Du et al.~\cite{du_gesture-_2022}, Aschenbrenner et al.~\cite{aschenbrenner_collaborative_2018}, and Hormaza et al.~\cite{hormaza_-line_2019} also explored the use of voice enhancement on the user side. In the environment, whether real or virtual, a few studies employed live video enhancement, opening a live video window within the environment, such as Zinchenko et al.'s work~\cite{zinchenko_autonomous_2021}.

\subsection{Multi-player and Multi-Robot Interaction Support}
\label{sec:multi}

\begin{figure}[htb]
  \centering
  \includegraphics[width=1\linewidth]{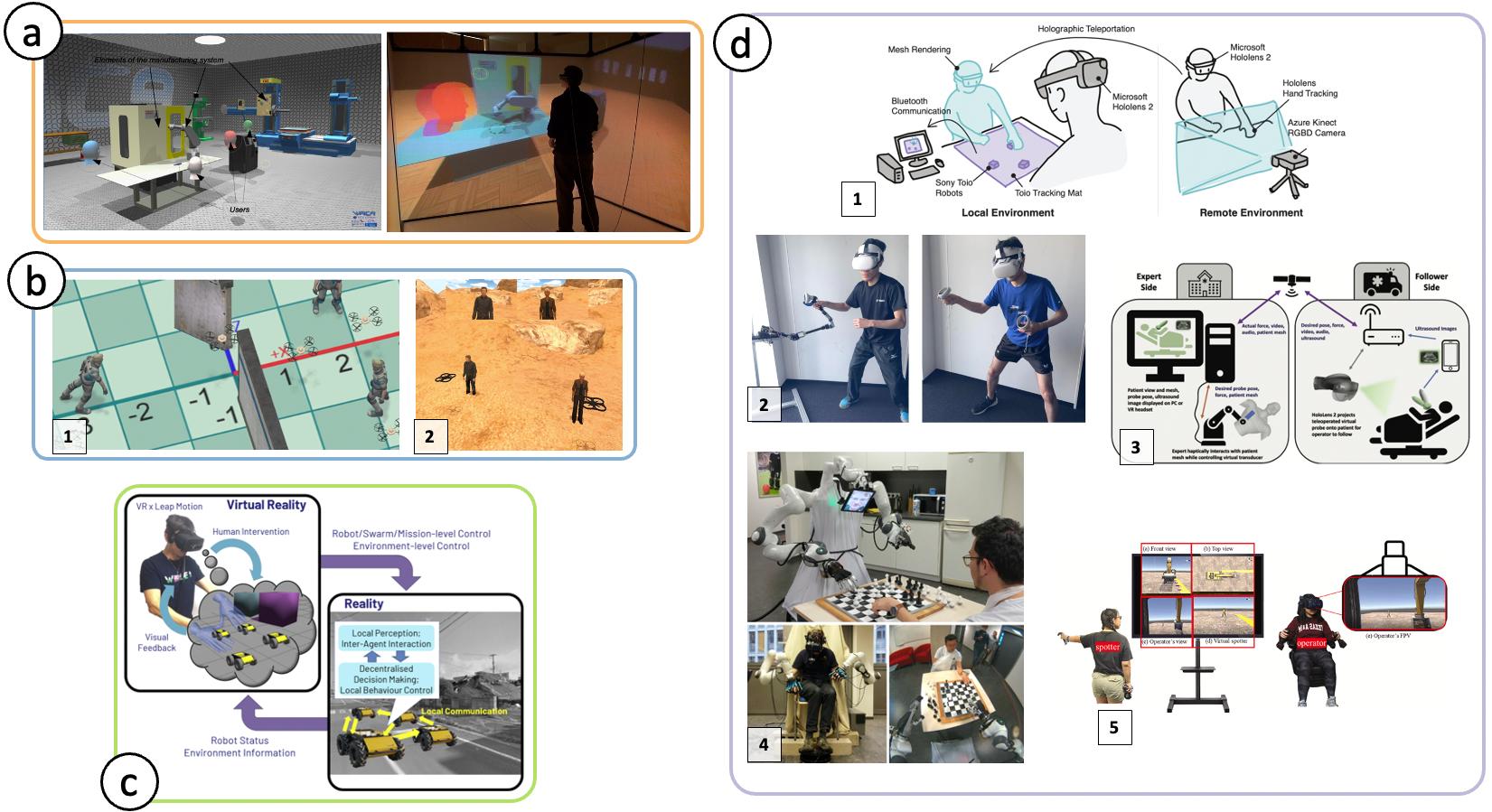}
  \caption{Examples of systems that support remote multiplayer or multi-robot interaction: a) Multi-user remote operate one robot~\cite{galambos_design_2015}; b) Multi-user remote operate multi-robot, 1: Multi-user control of multiple drones~\cite{phan_mixed_2018}, 2: Two users control two drones~\cite{honig_mixed_2015}; c) One user remote operate multi-robot~\cite{jang_omnipotent_2021}; d) A local user and a remote user cooperate together to control a robot, 1: Local and remote users collaborate with desktop robots~\cite{ihara2023holobots}, 2: Coach and learner remotely exercise table tennis in VR via robotic arm~\cite{fuchino2023t2remoter}, 3: Experts remotely guide novices in using medical equipment~\cite{black2023human}, 4: The local user controls the robot to play chess with the remote user~\cite{schwarz_nimbro_2021}, 5: The conductor remotely directs the user to maneuver the excavator~\cite{liu2023multi}.}
  \label{fig:multi}
\end{figure}

In our analysis, 81\% of the included papers did not support multi-player or multi-robot interaction, and all of their collaboration involved one user or operator collaborating with a single robot. Only 19\% of the interaction paradigms described in the papers supported multi-player or multi-robot interaction. One of the most common interaction paradigms involved one user interacting remotely with a single robot, while another user was situated next to the remote robot (N = 12). For instance, papers by Mourtzis et al.~\cite{mourtzis_augmented_2017,mourtzis_augmented_2019}, Black et al.~\cite{black2023human}, Fuchino et al.~\cite{fuchino2023t2remoter} and Moniri et al.~\cite{moniri_human_2016} propose that a novice operator or worker on the robot side receives instruction from a remote expert or commander using XR for environmental awareness and communication. 
The collaboration model in both Jang et al.~\cite{jang_omnipotent_2021} and Gong et al.'s~\cite{gong_real-time_2017} studies involved one user collaborating with multiple robots (N = 2). The remaining modes include multiple users interacting with multiple robots, such as the work of Phan et al.~\cite{phan_mixed_2018} and Honing et al.~\cite{honig_mixed_2015} (N = 2), multiple users operating a single robot, as exemplified by Galambos et al.~\cite{galambos_design_2015} (N = 1), and one user collaborating with multiple robots while multiple users are present on the robot side, as demonstrated by Aschenbrenner et al.~\cite{aschenbrenner_collaborative_2018} and Walker et al.~\cite{walker_mixed_2021} (N = 2).

\subsection{Exploration of System Latency Issues}
\label{sec:latencyResult}
\begin{figure}[htb]
  \centering
  \includegraphics[width=1\linewidth]{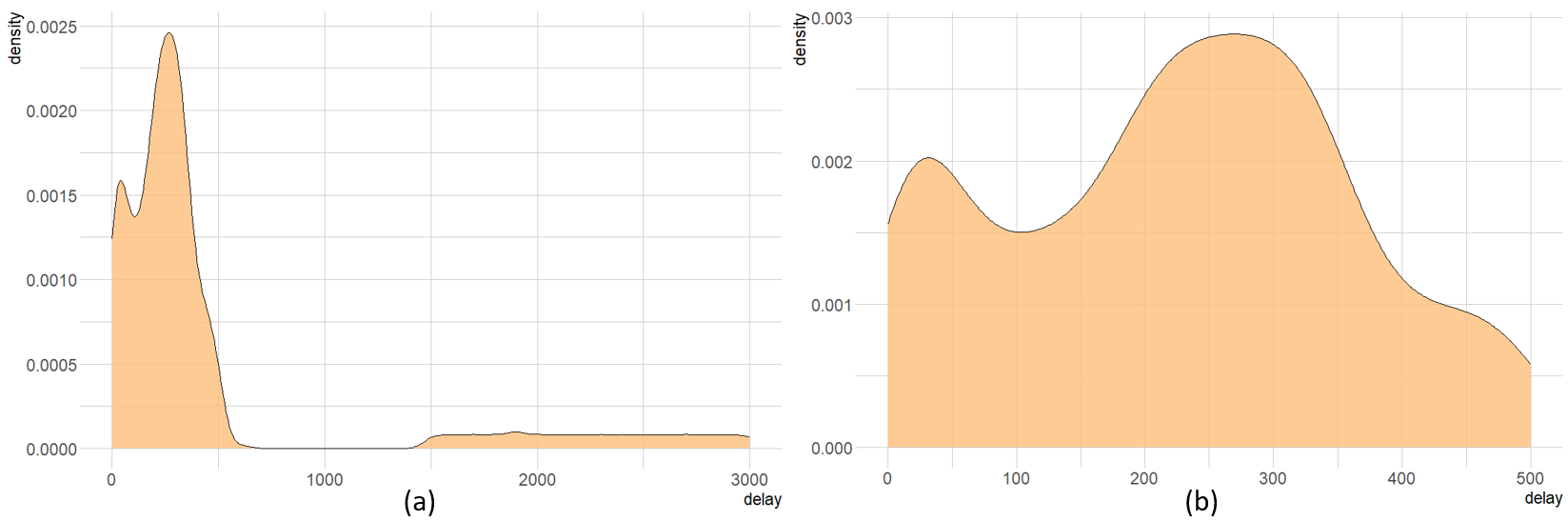}
  \caption{Density distribution of system latency times (unit: ms) in the included studies. (a) Latency times were reported by all included studies; (b) Latency was scaled to 0 -- 500ms to show more detail.}
  \label{fig:density}
\end{figure}

In our data extraction, we discovered that system latency effects are not considered in nearly half of the studies (49\%), despite the fact that remote control latency duration is a critical factor influencing user experience and task accuracy. A significant number of our included studies simply acknowledged the presence of system latency or claimed that their systems experienced delays, without providing specific measurements or quantifying the duration of their system latency (23\%). Only 28\% of the studies explicitly analyzed the latency of the system, and we compiled the reported study times in Figure~\ref{fig:density}. Figure~\ref{fig:density} (a) displays the latency times reported by all studies, revealing that the majority of studies had latency times within the range of less than $500 ms$, and only a few studies had latency times in the range of greater than $1000 ms$. Focusing on the latency times less than $500 ms$, as seen in Figure~\ref{fig:density} (b), we found that most studies had delay times within the range of $200 ms$ to $400 ms$. Among the included studies, several stand out for their unique approaches. Le et al.'s~\cite{le_intuitive_2020} research focused on controlling system latency and comparing the impact of different latencies on user experience. In contrast, McHenry et al.'s~\cite{mchenry_predictive_2021} study employed an asynchronous system operation to overcome the high latency associated with Earth-Moon transmission. These studies highlight distinct strategies for addressing latency in extended reality systems for remote robotic control.

\section{Discussion}
\label{sec:discussion}
\begin{figure}[htp]
  \centering
  \includegraphics[width = 1.3\textwidth, angle=90]{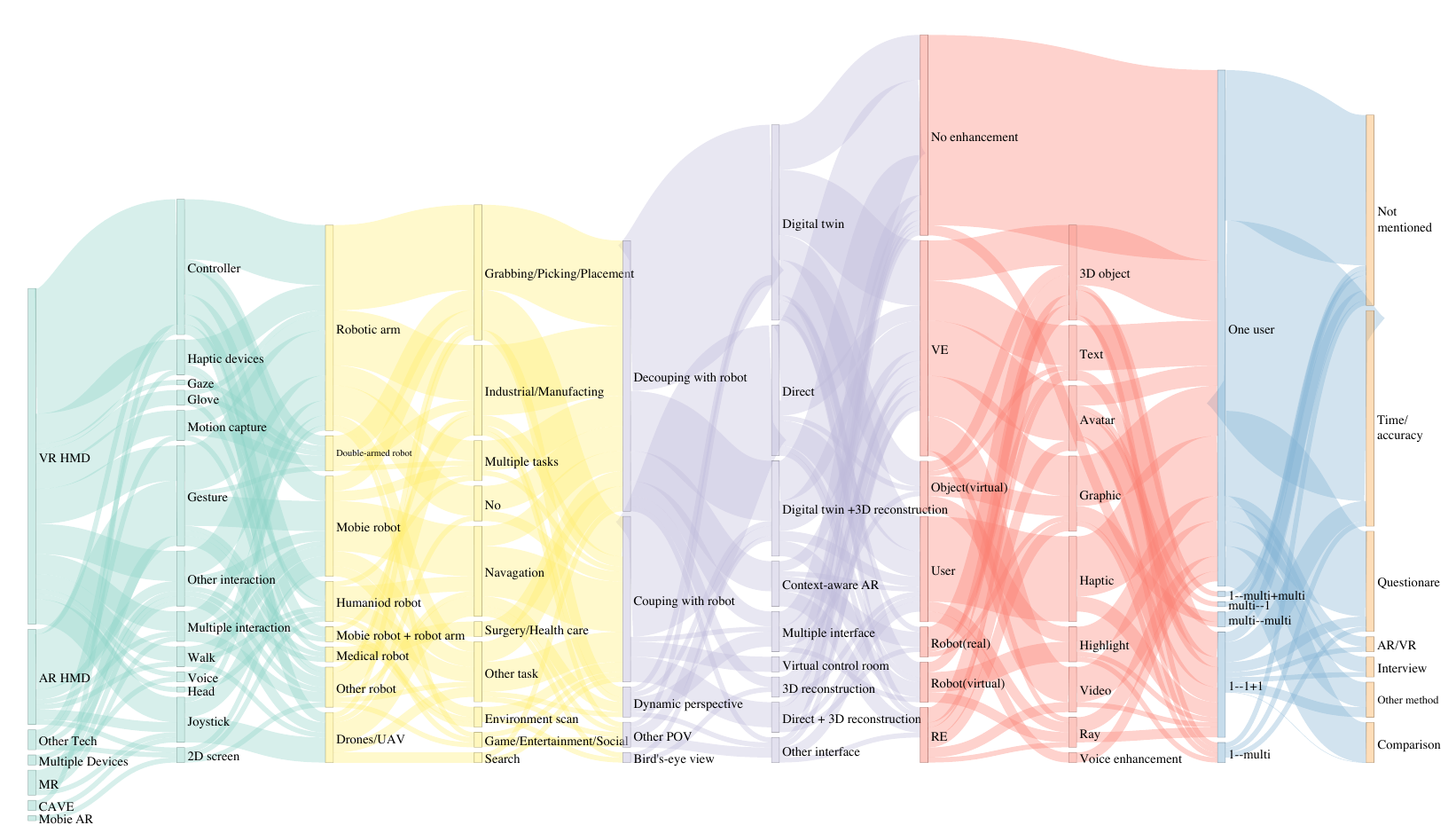}
  \caption{Sankey Diagram, 
  a visualization with overall counts of characteristics across all dimensions. Different stages (nodes) colors are used to distinguish different dimensions; \fcolorbox{white}{greenxian}{\textcolor{greenxian}{a}}: XR technologies and interaction modalities; \fcolorbox{white}{yellowxian}{\textcolor{yellowxian}{a}}: Robot and specifictasks classification; \fcolorbox{white}{purplexian}{\textcolor{purplexian}{a}}: User's Perspective and virtual interface; \fcolorbox{white}{redxian}{\textcolor{redxian}{a}}: Enhancement locations and types; \fcolorbox{white}{bluexian}{\textcolor{bluexian}{a}}: Support for multi-user/multi-robot; \fcolorbox{white}{orangexian}{\textcolor{orangexian}{a}}: Evaluation of tasks. Where the width of the flow (links) is proportional to the total number of articles we reviewed (N = 100).}
  \label{fig:sankey}
\end{figure}

Based on our data extraction results, Figure~\ref{fig:sankey} provides a visual summary of the number of papers associated with each dimension explored in this survey article. It is important to note that the prior section depicts the factual aspects of the collected articles. Next, we have to interpret these facts and highlight the insights. Thus, this section delves into the prevalent strategies and discrepancies observed across these selected dimensional features. Our discussion aims to shed light on the critical aspects of human-robot interaction and XR technologies, highlighting the areas of convergence and divergence among the studies reviewed. By examining these patterns, we seek to identify opportunities for future research and development in this rapidly evolving field, ultimately contributing to more effective and seamless collaboration among humans and robots.

\subsection{Impact of different robot types on remote HRI}

\subsubsection{How different robot types influence user interface and user perspective.}

\begin{figure}[htp]
  \centering
  \includegraphics[width = 1\textwidth]{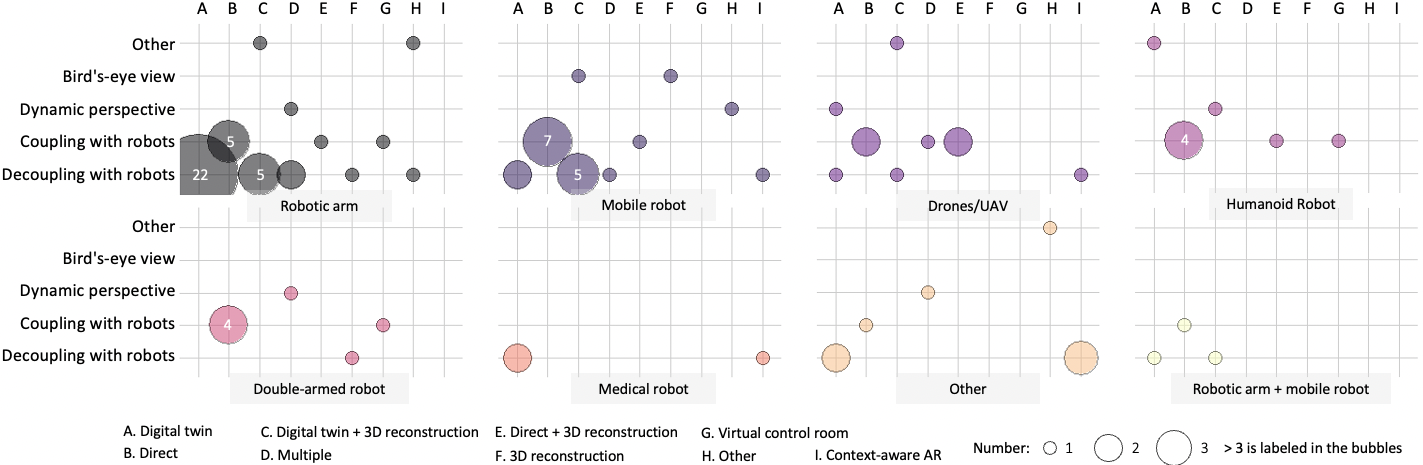}
  \caption{Bubble diagram of the relationship between types of robots and user interface and perspective. Different colored bubbles represent different robot types, the vertical axis indicates different user perspectives, and the horizontal axes A to I indicate different user interfaces. The size of the bubbles is proportional to the number of articles; the bigger the bubble, the more articles in the corresponding category, or vice versa. e.g., 
among the 100 papers reviewed, one study utilized a robotic arm, adopting a perspective of coupling with the robot through a virtual control room interface, while 22 articles also focusing on robotic arms, employed a decoupling perspective using a digital twin user interface.}
  \label{fig:robotUIPov}
\end{figure}

The wide variety of robots, each with its distinct characteristics and capabilities, inherently affects the design choices for user interfaces and perspectives in remote human-robot interaction (HRI) systems. Our analysis has identified specific associations between various types of robots and their corresponding tasks (Section~\ref{sec:robotTypesTasks}), which are critical to shaping the user interface and perspective design (See Figure~\ref{fig:robotUIPov}).

As analyzed in Section~\ref{sec:robotTypesTasks}, industrial robots, such as robotic arms and double-armed robots, are primarily deployed for production tasks. Given their limited mobility~\cite{almurib2012review}, these robots necessitate a user interface and perspective that focuses primarily on precision control and manipulation~\cite{cutkosky2012robotic}, rather than navigation. For instance, user interfaces designed for double-armed robots and robotic arms often facilitate multi-viewpoint together with dynamic perspective observation to assist users~\cite{xu_shared-control_2022,zhou_intuitive_2020}. A key distinguishing factor between these two robot types is their preferred control method in their user interface designs: two-armed robots typically favour direct operation, whereas robotic arms generally reflect a digital twin~\cite{park_interactive_2018,su_mixed_2021,vagvolgyi_scene_2018} (i.e., the data mapping between the virtuality and physical worlds). This distinction could stem from the fact that the double-armed configuration of the robots aligns closely with the human dual-arm anatomy~\cite{girbes2020haptic}. This alignment facilitates direct control of the remote robot and couples the user's perspective with the robot, potentially reducing the user's learning curve while making the operation more intuitive.

On the contrary, mobile robots, drones, and certain humanoid robots with advanced mobility are designed for tasks that require navigation, search, and environmental scanning. These tasks require user interfaces and perspectives that enhance spatial awareness and promote dynamic movement within the remote environment~\cite{yasuda2020autonomous}. For instance, mobile robot interfaces uniquely support a \textcolor{black}{bird's-eye view}, assisting users in comprehending the remote three-dimensional environment~\cite{walker_mixed_2021,aschenbrenner_collaborative_2018}.

In addition to the above interface designs related to the type of task the robot is assigned, it can be noticed through the bubble diagram that overall, there are increasing numbers of designs using direct interfaces in all robot types except medical robots. Perhaps one reason is that the cost of such direct interfaces is minimal and does not require 3D reconstruction or the creation of a digital twin of the real robot~\cite{hou2021exploring,adamenko2020review}. Moreover, we find that such direct interfaces are often coupled with the robot perspective, except for being more intuitive, probably also for cost reduction considerations. This design often requires only one or two cameras mounted on the robot~\cite{zhang_extended_2018,gai_new_2020,bai_kinect-based_2018,sanfilippo2022mixed}.

Finally, humanoid robots, known for their distinct social attributes, are typically engaged in tasks related to human or social activities~\cite{fox2021relationship}. In such scenarios, user interface and perspective designs should be geared towards intuitive control and interaction, allowing users to interact with the robot more human-like and socially.

\subsubsection{Types of robots affect virtual enhancements.}

In our review, medical robots extensively utilize virtual enhancement elements, with all examined medical robots incorporating such enhancements in their interactions~\cite{zinchenko_autonomous_2021,trejo_user_2018,black2023human}. We acknowledge that this observation may be influenced by the limited number of medical robot studies (only 3) included in our review, which may not provide a sufficient representation of the field. Alternatively, this high utilization of virtual enhancements could be attributed to the unique requirements of medical task scenarios, which require full use of the XR capabilities to assist operators~\cite{taylor2016medical}, such as physicians. 

Robotic arms, an industrial robot type characterized by its rising popularity, also demonstrated a high use of virtual augmentation elements, with 61\% of the examined robotic arms incorporating one or more such elements. This prevalence may be due to the extensive research focus on XR-based remote control of robotic arms, leading to a greater exploration of XR's unique augmentation characteristics. In particular, robotic arms were the only robot type that augmented virtual objects, which could be associated with their common task of picking and placing objects~\cite{wongphati2012you}. This task may necessitate additional augmentation on objects to enhance user-manipulation capabilities. 

Furthermore, robotic arms and mobile robots were the only types that utilized text overlays in the virtual environment. This feature may also be related to the tasks performed by robotic arms and mobile robots. Robotic arms are often used in professional contexts and industrial environments, where remote operators may benefit from text prompts or reminders for the next operational task, as exemplified in the design by Wang et al.~\cite{wang_digital_2021,wang_virtual_2019,wang_modeling_2019}. The remote operator of the mobile robot can also get information about the orientation of the robot movement from the text prompts~\cite{szczurek2023multimodal,kanazawa2023considerations}.

The unique mobility properties of mobile robots also influenced the choice of virtual enhancements. Mobile robots make extensive use of enhanced design in their environments (64.29\%), both virtual and real. Many designs used 3D object enhancements in the environment~\cite{walker_mixed_2021,cruz_ulloa_mixed-reality_2022,jang_omnipotent_2021}. This feature may be necessary to provide spatial location cues for users navigating remote mobile robots using XR, a requirement that does not apply to other non-mobile robots~\cite{bark2014personal}. Trinitatova et al.'s design uses a robotic arm but still uses 3D object enhancement in the environment~\cite{trinitatova2023study}. Their purpose was to use 3D spheres to indicate the center of the manipulated part of the robotic arm, still to indicate positional information in space. This suggests that the enhancement of 3D objects in the environment is often associated with positional information.

\subsection{Designing remote HRI system with users and scenarios }

The influence of users and scenarios on remote HRI system design is a crucial aspect to consider when developing effective human-robot collaboration experiences. Depending on the users' expertise, background, and preferences, as well as specific task scenarios, the design of remote HRI systems may need to be adapted accordingly to ensure optimal performance and usability. For example, adjust the user perspective and virtual interface accordingly, or select different enhanced design elements appropriately (see Section~\ref{sec:InterfaceEnhancedResult}).

\subsubsection{Expertise level of users and its impact on interaction design.}

Expert users, such as engineers or technicians, may have more advanced skills and familiarity with robotic systems, allowing them to handle more complex interfaces and control mechanisms. On the other hand, novice users, such as non-specialist workers or first-time users, may require more intuitive and user-friendly interfaces that prioritize ease of use and learning over advanced functionality~\cite{steinmetz2018razer,weintrop2018evaluating}. In addition, some of the systems designed for beginners will support the function of multiple operators to support remote expert guidance~\cite{black2023human}. 


\subsubsection{Designing adaptive systems to cater to diversified scenarios.}

Different task scenarios may necessitate varying levels of detail and control in the interface design. In scenarios where high precision and accuracy are required, such as remote surgery or delicate manipulation tasks, the interface may need to emphasize fine-grained control and provide more comprehensive feedback to the user~\cite{zhang2017reliable}. Conversely, in less demanding tasks, such as simple navigation or object transport, the interface could be designed with a more streamlined approach, focusing on usability and efficiency~\cite{bowman2002survey}. In addition, specific scenarios may require unique interface elements or features that cater to the particular challenges or requirements of the task. For example, in search and rescue missions, the interface may benefit from incorporating real-time mapping and tracking capabilities to aid in robot navigation and localization within the environment~\cite{queralta2020collaborative}.


\subsection{The role of XR in facilitating remote HRI}

 By providing immersive and interactive experiences, XR technologies have the potential to bridge the gap between users and remote robotic systems, enabling more efficient and natural collaboration.

\subsubsection{Enhancing user perspective and understanding of remote environments.}

One key advantage of XR in remote HRI is creating highly realistic and accurate digital twins of both the remote environment and the robot itself (see Section~\ref{sec:VirtualInterfaceResult}, digital twins are the most commonly used user interface). Users can better understand the remote workspace by simulating the robot's movements and actions within a virtual environment, making planning and executing tasks easier. Furthermore, the digital twin approach allows for more intuitive control mechanisms, as users can directly interact with the virtual representation of the robot, which in turn is mapped to the physical robot's behavior~\cite{schluse2018experimentable}. Another important aspect of XR in remote HRI is the provision of enhanced user perspectives (Section~\ref{sec:VirtualInterfaceResult}). XR technologies enable a wide range of viewing options, such as coupling the user's view with the robot, decoupling the view from the robot, or providing dynamic and flexible perspectives. These different perspectives allow users to adapt their viewing experience according to their preferences and specific task requirements, improving situational awareness and overall task performance~\cite{cockburn2009review}.



\subsubsection{Supporting multi-player or multi-robot interactions through XR.}
\label{sec:MultiDis}
One of the key benefits of XR in supporting multi-player or multi-robot interactions is the ability to create shared virtual spaces where users can collaborate and communicate more effectively~\cite{schroeder2010being}. By simulating multiple users or robots within a virtual environment, XR allows for better situational awareness and coordination, which is crucial for complex tasks that require teamwork and cooperation. This shared virtual space enables users to visualize the actions and intentions of other users or robots, improving overall task performance and efficiency. Moreover, XR technologies can be used to develop advanced user interfaces that cater to the unique requirements of multi-player or multi-robot scenarios. For instance, XR interfaces can provide real-time status updates, task assignments, and performance metrics for each user or robot, enabling better monitoring and decision-making. Additionally, XR interfaces can facilitate more intuitive control mechanisms, allowing users to seamlessly switch between controlling multiple robots or collaborating with other users. 

Despite the potential benefits of XR in multi-player or multi-robot interactions, our survey indicates that a significant portion of the included studies did not support such interactions (Section~\ref{sec:multi}), with most focusing on one user collaborating with one robot. However, several studies have explored various interaction paradigms, such as collaboration between a local user and a remote user\cite{ihara2023holobots,fuchino2023t2remoter,black2023human,liu2023multi,schwarz_nimbro_2021}, one user interacting with multiple robots~\cite{jang_omnipotent_2021}, multiple users operating a single robot~\cite{galambos_design_2015}, or multiple users collaborating with multiple robots~\cite{honig_mixed_2015,phan_mixed_2018}. These studies demonstrate the feasibility and potential advantages of leveraging XR technologies in multi-player or multi-robot HRI scenarios.

In addition to those discussed above, an important consideration in XR-enabled multi-player or multi-robot interaction systems is the challenge of latency, as emphasized by Jay et al.~\cite{jay2007modeling}, latency can severely impact the effectiveness of collaboration and the user's presence in the virtual environment, affecting the user's experience. High latency can disrupt coordination between users and robots, leading to errors and decreased task performance. Minimizing latency to enhance user experience and ensure smooth real-time interactions is important. Our review was pleased to find that many studies have noted the problem of latency (Section~\ref{sec:latencyResult}); however, addressing latency is especially important for systems that require the participation of multiple users and robots. As the number of participating users and robots increases, the latency problem may become more pronounced (See Section~\ref{sec:multiFuture} for further details).

\subsubsection{The use of multimodal enhancement in XR to improve remote operations.}

Multimodal enhancements in XR, such as visual, auditory, and haptic feedback~\cite{lee2008assessing}, can significantly improve the user experience and task performance in remote operations~\cite{kobayashi2016towards}. Our results (Section~\ref{sec:EhancementResult}) provide a detailed summary of the locations and kinds of multimodal enhancements. These enhancements provide users with more intuitive and immersive ways of perceiving and interacting with the remote environment and the robots involved. By leveraging multimodal feedback, XR can help bridge the gap between the user and the remote workspace, leading to more efficient and accurate task execution~\cite{maddikunta2022industry}.

For example, visual enhancements, such as highlighting specific parts of a robot or virtual object, can draw the user's attention to important elements and provide context-aware information~\cite{biocca2006attention}. Auditory feedback, such as spatial audio or voice commands, can deliver crucial information to the user and facilitate natural communication with other users or the robotic system itself~\cite{rosati2013role}. Haptic feedback, enabled through devices such as haptic gloves or virtual fixtures, can provide users with a more tangible sense of touch, enhancing their perception of the remote environment and improving their ability to perform complex tasks~\cite{pacchierotti2014improving}.

Our results indicate that a significant portion of the studies incorporated multimodal enhancements in their XR systems, with a focus on augmenting the virtual environment, user, and virtual objects. However, there remains room for further exploration in terms of augmenting real robots, real environments, and other aspects of remote operations.






\section{Challenges and Future Directions}
\label{sec:future}

After reviewing our comprehensive survey, we found that significant progress has been made in the development of remote HRI based on XR technologies, but many barriers and development opportunities still exist. The following sections discuss our reflections in detail, providing ideas for future researchers and suggestions for future developers. We will organize them into four main sections -- challenges in the selection of evaluation methods, unleashing the potential of XR in remote HRI, user-centered system design, and longitudinal studies and real-world deployment. We hope to work together with future researchers to provide innovative solutions to these challenges.

\subsection{Challenges in the selection of evaluation methods}


Selecting effective evaluation methods is a key challenge for future researchers exploring XR-based remote human-computer interaction. Our review found that a small half (38\%) of the relevant studies did not indicate any evaluation methods they used (see Section~\ref{sec:EvaluationOfTask}). By analyzing existing studies, evaluation methods can be broadly classified into two main categories: 1) evaluating the \textbf{system efficiency} and 2) evaluating the \textbf{user's experience}.

To evaluate system efficiency, including metrics such as system latency and task completion efficiency, the selection of evaluation methods should take into account the specific types of problems that these systems need to solve. For example, if help is needed for industrial production and assembly tasks, it may be more important to test the latency of the system as well as the accuracy and the time to complete specific tasks~\cite{gungor2009industrial}. In contrast, digital twin operator interfaces (XR as a teleoperation interface extension) may be more concerned with whether the digital twin corresponds to the behavior and actions of the real robot, usually comparing trajectory coordinates~\cite{li2022ar,liu2022genetic}. XR as a solution may significantly affect the system latency and thus change the efficiency of teleoperation~\cite{akyildiz2022wireless}, which needs to be considered by future researchers.

In contrast, the evaluation of the user experience in XR-based HRI involves more subjective user factors such as user satisfaction and ease of use of the system. Assessing these factors can greatly benefit from qualitative methods, including structured interviews and the use of various scales to measure user experience. Future researchers should consider the specific context of use, the target user group, and the purpose of the evaluation to select the appropriate standardized scale. For example, researchers can choose the scale \textit{NASA-TLX}~\cite{hart1988development} to measure users' subjective workload of the system, the \textit{System Usability Scale}~\cite{brooke1996sus} can use to measure system usability, and \textit{User Experience Questionnaire}~\cite{laugwitz2008construction} can measure the user experience of interactive products. In addition, interviews can provide deeper insight into user evaluations oriented to user perceptions, that may not be captured by separate quantitative measures~\cite{patton2002qualitative}. 

\subsection{Unleashing the Potential of XR in Remote HRI}

In the realm of remote HRI, the significance of XR is unmistakable. It presents burgeoning opportunities to transform the remote HRI paradigm, fostering more immersive, intuitive, and efficient interaction systems. These XR-enabled systems are poised to cater to the intricate demands of remote interactions with a diversity of robot types. Nevertheless, a considerable portion of XR's potential remains unexplored within the current research context, thus delineating promising pathways for forthcoming scholarly exploration.

\subsubsection{Multi-modal Remote Enhancement in XR}

XR offers a distinctive platform for multi-modal engagement, fostering a richer and more intuitive user interaction experience. Although our review has highlighted several instances where multi-modal enhancements have been used effectively, considerable scope remains for exploring innovative and integrative approaches to multi-modal remote interaction within the XR framework. Future research should actively pursue the understanding of the synergy of multi-modal cues, such as visual, auditory, and haptic cues and their roles in enhancing user comprehension and control in robot teleoperation.

In the visual domain, considerations could revolve around the optimal selection of visual enhancements for distinct locations. For instance, questions might arise about the efficiency of a 3D object overlay in virtual environments versus highlighted cues within the user's field of view for more intuitive navigation of mobile robots. Similarly, object manipulation can be guided by highlighting or by using pictorial markers, such as arrows, for quicker responses from remote operators. In addition, different robots and tasks can affect the design of visual enhancements. Future research could focus on exploring and evaluating interface designs for different categories of robots and tasks.

The possibility of tailored remote enhancements for different robot types and task types presents an intriguing opportunity for research. It is plausible that auditory cues may need to be designed differently for various robots and tasks. Regarding haptic feedback, there may be requirements for additional haptic devices, and pseudo-haptics could serve as an alternative~\cite{ujitoko2021survey}. Introducing new haptic devices should be carefully weighed against the risk of inadvertently increasing the physical burden on remote operators~\cite{wang2022vibroweight}.

The reviewed literature did not sufficiently cover the comprehensive consideration of these multimodal design specifics, especially the integration of multiple modalities. Future research should thoroughly examine these multichannel interactions, providing insight into optimal strategies to reduce cognitive load, enhance task performance, and heighten user immersion.

\subsubsection{Multi-player and multi-robot interactions and system latency}
\label{sec:multiFuture}

XR is becoming more popular in industrial, collaborative, and social domains, and the need for systems that support multiple users and multiple robots is probably going to increase in the future~\cite{ziker2021cross}. Future researchers and developers are required to be aware of the impact of system latency, in addition to designing efficient and intuitive XR systems to cope with these potential demands(see Section~\ref{sec:MultiDis}). Waltemate et al.~\cite{waltemate2016impact} indicated that although latency is unavoidable in VR applications, latency above $75ms$ affects the perception of motor performance and simultaneity. Whereas a latency greater than $125ms$ decreases the user's sense of agency and body ownership, which is worsened by more than $300ms$. Although our survey found that the latency for most included studies was in a reasonable range -- $200ms$ to $400ms$, within that range, users may perceive delays in the system, but not enough to completely crash the user experience. With the addition of more people or robots, the latency effect may get worse. This is an important issue for future researchers to be aware of. However, Waltemate et al. also pointed out that whether participants notice latency in virtual environments may depend on the motor task and its performance rather than the physical latency~\cite{waltemate2016impact}. This suggests that a user-friendly operating design may largely compensate for the negative effects of latency. For operational tasks that require precise control, it may be necessary to control the delay to within $75ms$.

\subsubsection{Navigating Complex Environments with XR in Remote HRI}

In the domain of remote HRI, operating in complex environments presents significant challenges that could be substantially mitigated with the judicious application of XR technologies~\cite{li2018human}. Robots may need to navigate through areas with varying terrain, unpredictable obstacles, or dynamic conditions, all of which pose different sets of complexities for remote operators. The fidelity with which these environments can be reproduced in an XR interface could have a substantial impact on the effectiveness of the HRI.

At present, the majority of systems employ a static third-person perspective or are coupled with a robot to relay environmental data to remote users. However, this approach, whether it's a singular viewpoint or merely a replication of the robot's perspective, may not provide the user with sufficient context to make optimal decisions. Advanced XR techniques have the potential to offer a comprehensive environmental overview, such as the dynamic or bird's-eye view, as investigated in a handful of studies within our review. However, these investigations remain sparse.

Future research could concentrate on the integration of multiple perspectives~\cite{trafton2005enabling}, such as a primary view linked to the robot, a third-person perspective for observing the robot, and a top-down bird's-eye view. This multi-perspective approach could enhance remote users' comprehension of the distant environment. It might even be feasible to establish a virtual environment camera to monitor the user avatar's operational state and the robot's digital twin from a third-person perspective in the virtual environment~\cite{benford2001collaborative}, potentially mitigating risks associated with certain tasks.

Another promising avenue for future research is the augmentation of the environment through multimodality, such as incorporating haptic feedback and auditory cues to enrich the user's perception of the remote environment. Additionally, the concept of an adaptive environment reconstruction system presents a potential research direction. Different robots and tasks may necessitate varying degrees of environmental reconstruction fidelity. For instance, pick-and-place tasks may only require low-fidelity reconstruction of the operator's table, while geological exploration tasks may demand high-fidelity environmental reconstruction. Implementing adaptive environmental reconstruction tailored to specific tasks and robots could potentially reduce system latency and prevent bandwidth wastage~\cite{wien2007real}.

Leveraging XR's capacity to reconstruct remote environments could enhance remote operators' spatial awareness, thereby improving navigation and task performance. Future research should prioritize the development of advanced environmental reconstruction techniques to provide a more comprehensive and real-time depiction of complex environments.

\subsubsection{Digital Twin in XR-based Remote Human-Robot Interaction}


The development and utilization of digital twins in XR-based remote human-robot interaction (HRI) presents a range of opportunities for future research. In the studies we reviewed, digital twins have emerged as an important component in enhancing the interaction and integration between physical and virtual environments. Digital twins can be used as more intuitive interfaces in XR-based human-computer interaction to improve system efficiency and user experience, and are more widely used in a variety of industrial and social scenarios~\cite{schluse2018experimentable}.

Future research in this area should focus on improving the fidelity of digital twins and their real-time synchronization with their physical counterparts. This includes improving the accuracy with which digital twins simulate complex physical processes and dynamics, which is critical for applications in industries such as manufacturing, healthcare, and urban planning. Additionally, the integration of advanced machine learning and artificial intelligence techniques can provide smarter and more autonomous digital twins capable of predictive maintenance, adaptive learning, and decision support~\cite{huang2021survey}.

Digital twins not only enhance the interaction between physical and virtual environments, but also provide possibilities for the study of scenarios that are impossible or impractical to test in reality. For example, the design by Su et al.~\cite{su_mixed_2021} displays both the zoomed-in parts of the task's operational details and a scaled-down model of the robotic arm's digital twin from the user's perspective. Such a design allows the operator to observe both the local details and the overall motion of the robotic arm at the same time, which is not possible in reality. This design takes full advantage of the potential of XR and digital twins, and future designs could make efforts to explore XR and digital twin capabilities that cannot be realized in reality.

\subsection{User-centered System Design}

As the domain of XR-based remote robotics operating systems continues to mature, the necessity for user-centered system design becomes increasingly apparent. This design philosophy ensures the development of systems that strike a balance between technological sophistication and user accessibility, fostering an intuitive user experience.

A critical facet of user-centered design is the accommodation of diverse user proficiency levels. Future research should strive to engineer systems that are universally accessible and efficient, catering to users across the proficiency spectrum. This could entail the design of adaptive interfaces that calibrate in response to a user's skill level, or the provision of comprehensive training modules to facilitate user familiarization with the system. Furthermore, the unique needs and preferences of different user groups warrant consideration. For instance, a system intended for professional engineers might necessitate a focus on precision and advanced functionalities, whereas a system tailored for the general populace might emphasize ease of use and intuitive controls. Usage scenarios are another important consideration, for instance, a system deployed for remote surgical procedures would have different specifications compared to one utilized for remote assembly tasks. Future research should discern the distinct needs of various user groups and also gain a deeper understanding of these diverse usage scenarios and their unique requirements.

Lastly, the innovative potential of XR needs to be considered in system design. XR technologies harbor the potential to craft immersive and engaging experiences that can significantly enhance user satisfaction and efficiency. For example, with the rising popularity of home robots, users operating a home robot via XR might interact with an avatar that could be a small animal or even a human, rather than a traditional robot. Future research should probe into innovative methods of harnessing XR's potential to elevate the user experience. This could involve the exploration of novel interaction techniques, the development of immersive feedback systems, or the invention of unique visualization methods.

\subsection{Longitudinal Studies and Real-world Deployment}

Our survey provides evidence of researchers' exploration of XR-based remote human-robot interaction systems across various domains such as industrial manufacturing and medicine. However, these works have predominantly focused on evaluating system performance within controlled laboratory settings, rather than real-world practical applications. Real environments are characterized by increased complexity and a multitude of factors that can potentially impact task performance, factors that are not present in laboratory settings. For instance, the performance of a system operating in a real field environment may be influenced by natural elements like strong winds, which can be mitigated or eliminated altogether in laboratory settings. Moreover, laboratory evaluations typically entail short durations, whereas real-world industrial applications demand long-term performance. Neglecting consistent usage over time can escalate maintenance costs and diminish operational efficiency. In summary, the efficacy observed in XR-based remote human-robot interaction systems within laboratory settings may not necessarily translate to practical industrial applications.

Future research should focus on improving the assessment of systems in real-world settings. Initially, it is necessary to assess the system's adaptability by evaluating job completion performance in uncontrolled industrial environments. Furthermore, it is recommended that the experiment length be extended to days and months, in order to evaluate the system's long-term performance and user behaviors in the wild, therefore guaranteeing its reliability, usability and consistency.

\section{Conclusion}
\label{sec:conclusion}

This paper explores the application of Extended Reality (XR) technologies in the emerging field of remote Human-Robot Interaction (HRI), through a comprehensive review and in-depth analysis of relevant 100 literature. The purpose of this research is to shed light on the potential, obstacles, and future research directions of XR as applied to remote HRI.

Our findings emphasize the transformative role XR has assumed in reconceptualizing the remote HRI paradigm, thereby enabling interaction systems that are simultaneously more engaging, intuitive, and efficient. We have taken an analytical approach to delineate the influence exerted by different robotic types on user interfaces and perspectives and highlight the necessity of custom-tailored designs that properly consider the unique features of the robot and task requirements. Furthermore, we scrutinize the employment of multichannel enhancements as a potential strategy for supporting teleoperation, while emphasizing the significance of a user-centered design in system development. Our study also uncovers several pertinent areas that warrant further investigation, inclusive but not limited to facilitating multi-user or multi-robot interactions and addressing challenges associated with latency and real-time performance.

As it stands, the prevailing system design appears deficient in fully leveraging the capabilities offered by XR technology. There exists an urgent need for additional research dedicated to advancing user-friendly XR system design for remote HRI. The insights presented in this paper not only contribute to the ongoing evolution of remote HRI but also serve as an indispensable resource for researchers, practitioners, and system designers who aim to use XR technology to optimize human-computer interaction.

\bibliographystyle{ACM-Reference-Format}
\bibliography{mybib}

\appendix
\section{Data Extraction List for Included Publication}
\label{appendix}
\begin{table}[tbh!]
\caption{XR Technologies and Interaction Modalities (DE4 and DE5)}
\label{Appendix_Table_Tech}
\scriptsize
\resizebox{1\linewidth}{!}{ 
\begin{tabular}{p{80pt}p{400pt}}
\toprule
\multicolumn{1}{c}{Category}   & \multicolumn{1}{c}{Citations} \\ \midrule
{\color[HTML]{82C1DA} \textbf{XR technologies}}        &                                                       \\
VR HMD                                                 &\cite{chen_enhanced_2022}\cite{xu_design_2022}\cite{ponomareva_grasplook_2021}\cite{wei_multi-view_2021}\cite{vu_investigation_2022}\cite{lipton_baxters_2017}\cite{su_mixed_2021}\cite{walker_mixed_2021}\cite{wibowo_improving_2021}\cite{zein_deep_2021}\cite{vagvolgyi_scene_2018}\cite{kulikovskiy_can_2021}\cite{zhou_intuitive_2020}\cite{chen_development_2017}\cite{park_interactive_2018}\cite{stotko_vr_2019}\cite{peppoloni_augmented_2015}\cite{wang_virtual_2019}\cite{chen_real-time_2020}\cite{galambos_design_2015}\cite{kuo_development_2021}\cite{yun_immersive_2022}\cite{xie_framework_2022}\cite{schwarz_nimbro_2021}\cite{cichon_digital_2018}\cite{vempati_virtual_2019}\cite{fani_simplifying_2018}\cite{zinchenko_autonomous_2021}\cite{codd-downey_ros_2014}\cite{bai_kinect-based_2018}\cite{peppoloni_immersive_2015}\\
 &\cite{takacs_towards_2015}\cite{phan_mixed_2018}\cite{omarali_virtual_2020}\cite{xu_shared-control_2022}\cite{jang_omnipotent_2021}\cite{jang_virtual_2021}\cite{wang_human-centered_2019}\cite{kalinov_warevr_2021}\cite{trejo_user_2018}\cite{theofilis_panoramic_2016}\cite{zhao_robot_2017}\cite{moniri_human_2016}\cite{xu_novel_2018}\cite{le_intuitive_2020}\cite{sun_haptic-feedback_2021}\cite{cousins_development_2017}\cite{wang_digital_2021}\cite{hormaza_-line_2019}\cite{brizzi_effects_2018}\cite{grzeskowiak_toward_2020}\cite{wang_novel_2017}\cite{bian_interface_2018}\cite{wang_modeling_2019}\cite{zhang_extended_2018}\cite{nagy_towards_2019}\cite{mchenry_predictive_2021}\cite{aschenbrenner_collaborative_2018}\cite{sakashita2023vroxy}\cite{nandhini2023teleoperation}\cite{kim2023towards}\cite{fan2023digital}\cite{abdulsalam2023vitrob}\\
  &\cite{kanazawa2023considerations}\cite{fuchino2023t2remoter}\cite{trinitatova2023study}\cite{nenna2023enhanced}\cite{xia2023visual}\cite{liu2023multi}\cite{sanfilippo2022mixed}\\
AR HMD                                                 &\cite{mourtzis_augmented_2019}\cite{hedayati_improving_2018}\cite{mourtzis_augmented_2017}\cite{xie_framework_2022}\cite{hernandez_increasing_2020}\cite{gai_new_2020}\cite{du_gesture-_2022}\cite{walker_robot_2019}\cite{xue_er_shamaine_rostar_2020}\cite{du_novel_2019}\cite{choi_xr-based_2022}\cite{gong_real-time_2017}\cite{xue_enabling_2020}\cite{aschenbrenner_collaborative_2018}\cite{szczurek2023multimodal}\cite{steinke2023future}\cite{meng2023virtual}\cite{rivera2023toward}\cite{black2023human}\cite{du2022intelligent}                                                       \\
MR                                                     &\cite{cruz_ulloa_mixed-reality_2022}\cite{ai_real-time_2016}\cite{honig_mixed_2015}\cite{ihara2023holobots}\cite{zaman2023vicarious}                                                       \\
CAVE                                                   &\cite{galambos_design_2015}\cite{gammieri_coupling_2017}\cite{betancourt_exocentric_2022}                                                       \\
Mobie AR                                               &\cite{chacko_augmented_2020}                                                       \\
Other                                                  &\cite{stedman_vrtab-map_2022}\cite{rastogi_control_2019}\cite{yew_immersive_2017}\cite{mu_design_2021}                                                       \\ 
{\color[HTML]{82C1DA} \textbf{Interaction modalities}} &                                                       \\
Controller                                             &\cite{chen_enhanced_2022}\cite{wei_multi-view_2021}\cite{lipton_baxters_2017}\cite{walker_mixed_2021}\cite{kulikovskiy_can_2021}\cite{zhou_intuitive_2020}\cite{wang_virtual_2019}\cite{chen_real-time_2020}\cite{gammieri_coupling_2017}\cite{kuo_development_2021}\cite{vempati_virtual_2019}\cite{codd-downey_ros_2014}\cite{takacs_towards_2015}\cite{omarali_virtual_2020}\cite{xu_shared-control_2022}\cite{kalinov_warevr_2021}\cite{hernandez_increasing_2020}\cite{le_intuitive_2020}\cite{wang_novel_2017}\cite{nagy_towards_2019}\cite{nandhini2023teleoperation}\cite{abdulsalam2023vitrob}\cite{kanazawa2023considerations}\cite{fuchino2023t2remoter}\cite{nenna2023enhanced}\cite{sanfilippo2022mixed}\cite{yun_immersive_2022}\cite{zaman2023vicarious}\cite{fan2023digital}\cite{liu2023multi}\\
Gesture                                                &\cite{cruz_ulloa_mixed-reality_2022}\cite{mourtzis_augmented_2017}\cite{xie_framework_2022}\cite{peppoloni_augmented_2015}\cite{jang_omnipotent_2021}\cite{jang_virtual_2021}\cite{zhao_robot_2017}\cite{xu_novel_2018}\cite{cousins_development_2017}\cite{xue_er_shamaine_rostar_2020}\cite{bian_interface_2018}\cite{choi_xr-based_2022}\cite{xue_enabling_2020}\cite{ihara2023holobots}\cite{szczurek2023multimodal}\cite{meng2023virtual}\cite{black2023human}\cite{du_gesture-_2022}\cite{sakashita2023vroxy}\cite{zaman2023vicarious}\cite{rivera2023toward}                                                       \\
Joystick                                               &\cite{vu_investigation_2022}\cite{wibowo_improving_2021}\cite{zein_deep_2021}\cite{hedayati_improving_2018}\cite{stotko_vr_2019}\cite{ai_real-time_2016}\cite{betancourt_exocentric_2022}\cite{walker_robot_2019}\cite{xia2023visual}\\
Haptic devices                                         &\cite{ponomareva_grasplook_2021}\cite{chen_development_2017}\cite{su_mixed_2021}\cite{rastogi_control_2019}\cite{schwarz_nimbro_2021}\cite{trejo_user_2018}\cite{kim2023towards}\cite{yun_immersive_2022}\cite{fan2023digital}                                                       \\
Motion capture                                         &\cite{xu_design_2022}\cite{peppoloni_augmented_2015}\cite{fani_simplifying_2018}\cite{bai_kinect-based_2018}\cite{brizzi_effects_2018}                                                       \\
Walk                                                   &\cite{phan_mixed_2018}\cite{gai_new_2020}\cite{grzeskowiak_toward_2020}\cite{du_novel_2019}\cite{sakashita2023vroxy}                                                       \\
2D screen                                              &\cite{mu_design_2021}\cite{chacko_augmented_2020}\cite{gong_real-time_2017}\\
Voice                                                  &\cite{hormaza_-line_2019}\cite{aschenbrenner_collaborative_2018}\cite{rivera2023toward}                                                       \\
Glove                                                  &\cite{park_interactive_2018}\cite{sun_haptic-feedback_2021}\cite{trinitatova2023study}                                                       \\
Head                                                   &\cite{theofilis_panoramic_2016}\cite{sakashita2023vroxy}                                                       \\
Gaze                                                   &\cite{moniri_human_2016}                                                      \\
Other                                                  &\cite{stedman_vrtab-map_2022}\cite{mourtzis_augmented_2019}\cite{vagvolgyi_scene_2018}\cite{galambos_design_2015}\cite{cichon_digital_2018}\cite{zinchenko_autonomous_2021}\cite{wang_human-centered_2019}\cite{honig_mixed_2015}\cite{wang_digital_2021}\cite{du_novel_2019}\cite{mchenry_predictive_2021}\cite{liu2023multi}\\\bottomrule                                                      
\end{tabular}}
\end{table}

\begin{table}[tbh!]
\caption{Virtual Interface and the User's Perspective (DE6 and DE7)}
\label{Appendix_Table_Interface}
\scriptsize
\resizebox{1\linewidth}{!}{ 
\begin{tabular}{p{90pt}p{390pt}}
\toprule
\multicolumn{1}{c}{Category}   & \multicolumn{1}{c}{Citations} \\ \midrule
{\color[HTML]{82C1DA} \textbf{XR technologies}}        &                                                       \\
Digital twin &\cite{cruz_ulloa_mixed-reality_2022}\cite{su_mixed_2021}\cite{vagvolgyi_scene_2018}\cite{wang_virtual_2019}\cite{yew_immersive_2017}\cite{gammieri_coupling_2017}\cite{yun_immersive_2022}\cite{xie_framework_2022}\cite{zinchenko_autonomous_2021}\cite{peppoloni_augmented_2015}\cite{takacs_towards_2015}\cite{jang_virtual_2021}\cite{trejo_user_2018}\cite{betancourt_exocentric_2022}\cite{le_intuitive_2020}\cite{wang_digital_2021}\cite{du_gesture-_2022}\cite{walker_robot_2019}\cite{brizzi_effects_2018}\cite{xue_er_shamaine_rostar_2020}\cite{grzeskowiak_toward_2020}\cite{wang_modeling_2019}\cite{nagy_towards_2019}\cite{choi_xr-based_2022}\cite{mchenry_predictive_2021}\cite{xue_enabling_2020}\cite{nandhini2023teleoperation}\cite{fan2023digital}\cite{abdulsalam2023vitrob}\cite{trinitatova2023study}\cite{meng2023virtual}\\ 
&\cite{nenna2023enhanced}\cite{rivera2023toward} \\
Direct & \cite{xu_design_2022}\cite{wibowo_improving_2021}\cite{hedayati_improving_2018}\cite{hedayati_improving_2018}\cite{chen_development_2017}\cite{park_interactive_2018}\cite{stotko_vr_2019}\cite{peppoloni_augmented_2015}\cite{chen_real-time_2020}\cite{ai_real-time_2016}\cite{schwarz_nimbro_2021}\cite{fani_simplifying_2018}\cite{bai_kinect-based_2018}\cite{wang_human-centered_2019}\cite{mu_design_2021}\cite{theofilis_panoramic_2016}\cite{zhao_robot_2017}\cite{gai_new_2020}\cite{cousins_development_2017}\cite{chacko_augmented_2020}\cite{brizzi_effects_2018}\cite{bian_interface_2018}\cite{zhang_extended_2018}\cite{gong_real-time_2017}\cite{ihara2023holobots}\cite{zaman2023vicarious}\cite{xia2023visual}\cite{black2023human}\cite{sanfilippo2022mixed}\\
Digital twin+3D reconstruction &\cite{ponomareva_grasplook_2021}\cite{stedman_vrtab-map_2022}\cite{rastogi_control_2019}\cite{galambos_design_2015}\cite{kuo_development_2021}\cite{cichon_digital_2018}\cite{codd-downey_ros_2014}\cite{phan_mixed_2018}\cite{omarali_virtual_2020}\cite{jang_omnipotent_2021}\cite{honig_mixed_2015}\cite{wang_novel_2017}\cite{szczurek2023multimodal}\cite{du_gesture-_2022}\cite{aschenbrenner_collaborative_2018} \\
Multiple &\cite{vu_investigation_2022}\cite{zhou_intuitive_2020}\cite{xu_shared-control_2022}\cite{du_novel_2019}\cite{kanazawa2023considerations}\cite{liu2023multi} \\
Direct+3D reconstruction & \cite{chen_enhanced_2022}\cite{zein_deep_2021}\cite{vempati_virtual_2019}\cite{sakashita2023vroxy}\cite{kim2023towards}\\
3D reconstruction & \cite{wei_multi-view_2021}\cite{walker_mixed_2021}\cite{moniri_human_2016}\\
Virtual control room &\cite{lipton_baxters_2017}\cite{kulikovskiy_can_2021}\cite{steinke2023future}\\
Other & \cite{mourtzis_augmented_2019}\cite{mourtzis_augmented_2017}\cite{hernandez_increasing_2020}\cite{sun_haptic-feedback_2021}\cite{xu_novel_2018}\cite{hormaza_-line_2019}\cite{fuchino2023t2remoter}\\

{\color[HTML]{82C1DA} \textbf{User's perspective}}        &                                                       \\
Decoupling with robots &\cite{ponomareva_grasplook_2021}\cite{vu_investigation_2022}\cite{stedman_vrtab-map_2022}\cite{cruz_ulloa_mixed-reality_2022}\cite{mourtzis_augmented_2019}\cite{su_mixed_2021}\cite{hedayati_improving_2018}\cite{vagvolgyi_scene_2018}\cite{wang_virtual_2019}\cite{rastogi_control_2019}\cite{yew_immersive_2017}\cite{galambos_design_2015}\cite{gammieri_coupling_2017}\cite{kuo_development_2021}\cite{yun_immersive_2022}\cite{xie_framework_2022}\cite{cichon_digital_2018}\cite{zinchenko_autonomous_2021}\cite{codd-downey_ros_2014}\cite{peppoloni_immersive_2015}\cite{omarali_virtual_2020}\cite{jang_virtual_2021}\cite{trejo_user_2018}\cite{moniri_human_2016}\cite{xu_novel_2018}\cite{le_intuitive_2020}\cite{honig_mixed_2015}\cite{wang_digital_2021}\cite{du_gesture-_2022}\cite{hormaza_-line_2019}\cite{walker_robot_2019}\\
&\cite{xue_er_shamaine_rostar_2020}\cite{wang_novel_2017}\cite{wang_modeling_2019}\cite{du_novel_2019}\cite{nagy_towards_2019}\cite{choi_xr-based_2022}\cite{mchenry_predictive_2021}\cite{xue_enabling_2020}\cite{ihara2023holobots}\cite{zaman2023vicarious}\cite{nandhini2023teleoperation}\cite{szczurek2023multimodal}\cite{fan2023digital}\cite{abdulsalam2023vitrob}\cite{kanazawa2023considerations}\cite{meng2023virtual}\cite{nenna2023enhanced}\cite{rivera2023toward}\cite{du2022intelligent}\\
Coupling with robots & \cite{xu_design_2022}\cite{lipton_baxters_2017}\cite{wibowo_improving_2021}\cite{zein_deep_2021}\cite{kulikovskiy_can_2021}\cite{chen_development_2017}\cite{park_interactive_2018}\cite{stotko_vr_2019}\cite{peppoloni_augmented_2015}\cite{mourtzis_augmented_2017}\cite{chen_real-time_2020}\cite{ai_real-time_2016}\cite{schwarz_nimbro_2021}\cite{vempati_virtual_2019}\cite{fani_simplifying_2018}\cite{bai_kinect-based_2018}\cite{wang_human-centered_2019}\cite{mu_design_2021}\cite{kalinov_warevr_2021}\cite{theofilis_panoramic_2016}\cite{zhao_robot_2017}\cite{gai_new_2020}\cite{sun_haptic-feedback_2021}\cite{cousins_development_2017}\cite{chacko_augmented_2020}\cite{brizzi_effects_2018}\cite{bian_interface_2018}\cite{zhang_extended_2018}\cite{gong_real-time_2017}\cite{sakashita2023vroxy}\cite{kim2023towards}\cite{steinke2023future}\\
&\cite{xia2023visual}\cite{sanfilippo2022mixed}\\
Dynamic perspective & \cite{zhou_intuitive_2020}\cite{xu_shared-control_2022}\cite{hernandez_increasing_2020}\cite{betancourt_exocentric_2022}\cite{liu2023multi}\\
God perspective & \cite{walker_mixed_2021}\cite{jang_omnipotent_2021}\\
Other & \cite{chen_enhanced_2022}\cite{vu_investigation_2022}\cite{takacs_towards_2015}\cite{phan_mixed_2018}\cite{grzeskowiak_toward_2020}\cite{aschenbrenner_collaborative_2018}\cite{fuchino2023t2remoter}\cite{black2023human}\\\bottomrule
\end{tabular}}
\end{table}

\begin{table}[tbh!]
\caption{Robot Types and Specific Tasks Classification (DE8 and DE9)}
\label{Appendix_Table_Robot}
\scriptsize
\resizebox{1\linewidth}{!}{ 
\begin{tabular}{p{90pt}p{390pt}}
\toprule
\multicolumn{1}{c}{Category}   & \multicolumn{1}{c}{Citations} \\ \midrule
{\color[HTML]{82C1DA} \textbf{Robot types}}        &                                                       \\
Robotic arm & \cite{xu_design_2022}\cite{ponomareva_grasplook_2021}\cite{wei_multi-view_2021}\cite{vu_investigation_2022}\cite{su_mixed_2021}\cite{vagvolgyi_scene_2018}\cite{chen_development_2017}\cite{wang_virtual_2019}\cite{rastogi_control_2019}\cite{galambos_design_2015}\cite{gammieri_coupling_2017}\cite{kuo_development_2021}\cite{yun_immersive_2022}\cite{fani_simplifying_2018}\cite{bai_kinect-based_2018}\cite{peppoloni_immersive_2015}\cite{omarali_virtual_2020}\cite{xu_shared-control_2022}\cite{jang_virtual_2021}\cite{xu_novel_2018}\cite{le_intuitive_2020}\cite{cousins_development_2017}\cite{wang_digital_2021}\cite{hormaza_-line_2019}\cite{xue_er_shamaine_rostar_2020}\cite{wang_modeling_2019}\cite{du_novel_2019}\cite{nagy_towards_2019}\cite{choi_xr-based_2022}\cite{mchenry_predictive_2021}\\
&\cite{xue_enabling_2020}\cite{aschenbrenner_collaborative_2018}\cite{nandhini2023teleoperation}\cite{kim2023towards}\cite{fan2023digital}\cite{steinke2023future}\cite{fuchino2023t2remoter}\cite{trinitatova2023study}\cite{meng2023virtual}\cite{nenna2023enhanced}\cite{rivera2023toward}\\
Mobile robot & \cite{stedman_vrtab-map_2022}\cite{cruz_ulloa_mixed-reality_2022}\cite{walker_mixed_2021}\cite{wibowo_improving_2021}\cite{stotko_vr_2019}\cite{cichon_digital_2018}\cite{codd-downey_ros_2014}\cite{jang_omnipotent_2021}\cite{hernandez_increasing_2020}\cite{zhao_robot_2017}\cite{gai_new_2020}\cite{du_gesture-_2022}\cite{chacko_augmented_2020}\cite{wang_novel_2017}\cite{zhang_extended_2018}\cite{ihara2023holobots}\cite{sakashita2023vroxy}\cite{szczurek2023multimodal}\cite{kanazawa2023considerations}\cite{xia2023visual}\\
Drones/UAV & \cite{zein_deep_2021}\cite{hedayati_improving_2018}\cite{ai_real-time_2016}\cite{vempati_virtual_2019}\cite{takacs_towards_2015}\cite{wang_human-centered_2019}\cite{kalinov_warevr_2021}\cite{betancourt_exocentric_2022}\cite{honig_mixed_2015}\cite{walker_robot_2019}\\
Humanoid robot & \cite{chen_enhanced_2022}\cite{kulikovskiy_can_2021}\cite{schwarz_nimbro_2021}\cite{takacs_towards_2015}\cite{theofilis_panoramic_2016}\cite{grzeskowiak_toward_2020}\cite{gong_real-time_2017}\cite{sanfilippo2022mixed}\\
Double-armed robot & \cite{lipton_baxters_2017}\cite{zhou_intuitive_2020}\cite{park_interactive_2018}\cite{peppoloni_augmented_2015}\cite{moniri_human_2016}\cite{brizzi_effects_2018}\cite{bian_interface_2018}\\
Medical Robotics & \cite{zinchenko_autonomous_2021}\cite{trejo_user_2018}\cite{black2023human}\\
Other & \cite{mourtzis_augmented_2019}\cite{mourtzis_augmented_2017}\cite{yew_immersive_2017}\cite{xie_framework_2022}\cite{mu_design_2021}\cite{sun_haptic-feedback_2021}\cite{zaman2023vicarious}\cite{liu2023multi}\\
Robotic arm+mobile robot & \cite{chen_real-time_2020}\cite{abdulsalam2023vitrob}\cite{du2022intelligent}\\

{\color[HTML]{82C1DA} \textbf{Specific tasks}}        &                                                       \\
Grabbing/Picking/Placement & \cite{xu_design_2022}\cite{ponomareva_grasplook_2021}\cite{wei_multi-view_2021}\cite{vu_investigation_2022}\cite{cruz_ulloa_mixed-reality_2022}\cite{chen_development_2017}\cite{peppoloni_augmented_2015}\cite{kuo_development_2021}\cite{yun_immersive_2022}\cite{bai_kinect-based_2018}\cite{omarali_virtual_2020}\cite{xu_shared-control_2022}\cite{jang_virtual_2021}\cite{hernandez_increasing_2020}\cite{moniri_human_2016}\cite{honig_mixed_2015}\cite{brizzi_effects_2018}\cite{bian_interface_2018}\cite{mchenry_predictive_2021}\cite{xue_enabling_2020}\cite{zaman2023vicarious}\cite{fan2023digital}\cite{steinke2023future}\cite{kanazawa2023considerations}\cite{meng2023virtual}\cite{nenna2023enhanced}\cite{sanfilippo2022mixed}\\
Navigation & \cite{chen_enhanced_2022}\cite{stedman_vrtab-map_2022}\cite{wibowo_improving_2021}\cite{stotko_vr_2019}\cite{ai_real-time_2016}\cite{codd-downey_ros_2014}\cite{phan_mixed_2018}\cite{jang_omnipotent_2021}\cite{mu_design_2021}\cite{theofilis_panoramic_2016}\cite{betancourt_exocentric_2022}\cite{gai_new_2020}\cite{chacko_augmented_2020}\cite{walker_robot_2019}\cite{grzeskowiak_toward_2020}\cite{wang_novel_2017}\cite{zhang_extended_2018}\cite{xia2023visual}\\
Industrial/Manufacturing & \cite{mourtzis_augmented_2019}\cite{lipton_baxters_2017}\cite{su_mixed_2021}\cite{vagvolgyi_scene_2018}\cite{zhou_intuitive_2020}\cite{mourtzis_augmented_2017}\cite{wang_virtual_2019}\cite{yew_immersive_2017}\cite{galambos_design_2015}\cite{xie_framework_2022}\cite{fani_simplifying_2018}\cite{xu_novel_2018}\cite{wang_digital_2021}\cite{wang_modeling_2019}\cite{nagy_towards_2019}\cite{aschenbrenner_collaborative_2018}\cite{rivera2023toward}\cite{du2022intelligent}\\
Multiple & \cite{park_interactive_2018}\cite{rastogi_control_2019}\cite{cichon_digital_2018}\cite{peppoloni_immersive_2015}\cite{du_gesture-_2022}\cite{du_novel_2019}\\
No & \cite{gammieri_coupling_2017}\cite{wang_human-centered_2019}\cite{cousins_development_2017}\cite{xue_er_shamaine_rostar_2020}\cite{choi_xr-based_2022}\cite{nandhini2023teleoperation}\cite{trinitatova2023study}\\
Environment scan & \cite{walker_mixed_2021}\cite{chen_real-time_2020}\cite{zhao_robot_2017}\\
Surgery/Healthcare & \cite{zinchenko_autonomous_2021}\cite{trejo_user_2018}\cite{black2023human}\\
Search & \cite{kalinov_warevr_2021}\\
Game/Entertainment/Social & \cite{schwarz_nimbro_2021}\cite{sun_haptic-feedback_2021}\cite{fuchino2023t2remoter}\\
Other & \cite{zein_deep_2021}\cite{kulikovskiy_can_2021}\cite{vempati_virtual_2019}\cite{takacs_towards_2015}\cite{le_intuitive_2020}\cite{gong_real-time_2017}\cite{ihara2023holobots}\cite{sakashita2023vroxy}\cite{kim2023towards}\cite{szczurek2023multimodal}\cite{abdulsalam2023vitrob}\cite{liu2023multi}\\\bottomrule

\end{tabular}}
\end{table}

\begin{table}[tbh!]
\caption{Enhancement locations and types (DE13 and DE14)}
\label{Appendix_Table_Enhancement}
\scriptsize
\resizebox{1\linewidth}{!}{ 
\begin{tabular}{p{80pt}p{400pt}}
\toprule
\multicolumn{1}{c}{Category}   & \multicolumn{1}{c}{Citations} \\ \midrule
{\color[HTML]{82C1DA} \textbf{Enhancement location}}        &                                                       \\
VE(Virtual environment) & \cite{cruz_ulloa_mixed-reality_2022}\cite{walker_mixed_2021}\cite{zein_deep_2021}\cite{wang_virtual_2019}\cite{galambos_design_2015}\cite{vempati_virtual_2019}\cite{takacs_towards_2015}\cite{trejo_user_2018}\cite{cousins_development_2017}\cite{honig_mixed_2015}\cite{wang_digital_2021}\cite{grzeskowiak_toward_2020}\cite{wang_modeling_2019}\cite{gong_real-time_2017}\cite{steinke2023future}\cite{kanazawa2023considerations}\cite{meng2023virtual}\cite{liu2023multi}\cite{kuo_development_2021}\cite{zinchenko_autonomous_2021}\cite{xu_shared-control_2022}\cite{jang_omnipotent_2021}\cite{moniri_human_2016}\cite{xue_er_shamaine_rostar_2020}\cite{brizzi_effects_2018}\cite{choi_xr-based_2022}\cite{sakashita2023vroxy}\cite{zaman2023vicarious}\cite{szczurek2023multimodal}\cite{fan2023digital}\cite{fuchino2023t2remoter}\\
&\cite{trinitatova2023study}\cite{rivera2023toward}\cite{black2023human}\\
User & \cite{ponomareva_grasplook_2021}\cite{chen_development_2017}\cite{park_interactive_2018}\cite{rastogi_control_2019}\cite{schwarz_nimbro_2021}\cite{fani_simplifying_2018}\cite{sun_haptic-feedback_2021}\cite{du_gesture-_2022}\cite{kim2023towards}\cite{xia2023visual}\cite{su_mixed_2021}\cite{yun_immersive_2022}\cite{moniri_human_2016}\cite{choi_xr-based_2022}\cite{fan2023digital}\cite{fuchino2023t2remoter}\cite{trinitatova2023study}\cite{rivera2023toward}\cite{black2023human}\\
Robot(Virtual) & \cite{betancourt_exocentric_2022}\cite{nagy_towards_2019}\cite{mchenry_predictive_2021}\cite{kuo_development_2021}\cite{xu_shared-control_2022}\cite{jang_omnipotent_2021}\cite{brizzi_effects_2018}\cite{szczurek2023multimodal}\\
Robot(Real) & \cite{mourtzis_augmented_2019}\cite{mourtzis_augmented_2017}\cite{hedayati_improving_2018}\cite{choi_xr-based_2022}\cite{sakashita2023vroxy}\\
RE(Real environment) & \cite{hernandez_increasing_2020}\cite{chacko_augmented_2020}\cite{walker_robot_2019}\cite{ihara2023holobots}\cite{hedayati_improving_2018}\cite{zaman2023vicarious}\\
Object(Virtual) & \cite{vu_investigation_2022}\cite{vagvolgyi_scene_2018}\cite{gammieri_coupling_2017}\cite{xue_enabling_2020}\cite{su_mixed_2021}\cite{yun_immersive_2022}\cite{zinchenko_autonomous_2021}\cite{xue_er_shamaine_rostar_2020}\cite{brizzi_effects_2018}\\
No & \cite{chen_enhanced_2022}\cite{xu_design_2022}\cite{wei_multi-view_2021}\cite{stedman_vrtab-map_2022}\cite{lipton_baxters_2017}\cite{wibowo_improving_2021}\cite{kulikovskiy_can_2021}\cite{zhou_intuitive_2020}\cite{stotko_vr_2019}\cite{peppoloni_augmented_2015}\cite{chen_real-time_2020}\cite{yew_immersive_2017}\cite{xie_framework_2022}\cite{ai_real-time_2016}\cite{cichon_digital_2018}\cite{codd-downey_ros_2014}\cite{bai_kinect-based_2018}\cite{peppoloni_immersive_2015}\cite{phan_mixed_2018}\cite{omarali_virtual_2020}\cite{jang_virtual_2021}\cite{wang_human-centered_2019}\cite{mu_design_2021}\cite{kalinov_warevr_2021}\cite{theofilis_panoramic_2016}\cite{zhao_robot_2017}\cite{gai_new_2020}\cite{xu_novel_2018}\cite{le_intuitive_2020}\cite{hormaza_-line_2019}\cite{wang_novel_2017}\\
&\cite{bian_interface_2018}\cite{du_novel_2019}\cite{zhang_extended_2018}\cite{aschenbrenner_collaborative_2018}\cite{nandhini2023teleoperation}\cite{abdulsalam2023vitrob}\cite{nenna2023enhanced}\cite{du2022intelligent}\cite{sanfilippo2022mixed}\\

{\color[HTML]{82C1DA} \textbf{Enhancement types}}        &                                                       \\
Voice & \cite{du_gesture-_2022}\cite{rivera2023toward}\\
Video & \cite{steinke2023future}\cite{meng2023virtual}\cite{hedayati_improving_2018}\cite{zinchenko_autonomous_2021}\cite{xu_shared-control_2022}\cite{zaman2023vicarious}\cite{szczurek2023multimodal}\cite{fan2023digital}\cite{kanazawa2023considerations}\\
Text & \cite{wang_virtual_2019}\cite{wang_digital_2021}\cite{wang_modeling_2019}\cite{mourtzis_augmented_2019}\cite{kuo_development_2021}\cite{yun_immersive_2022}\cite{xu_shared-control_2022}\cite{chacko_augmented_2020}\cite{xue_er_shamaine_rostar_2020}\cite{szczurek2023multimodal}\cite{kanazawa2023considerations}\\
Ray & \cite{kuo_development_2021}\cite{moniri_human_2016}\cite{brizzi_effects_2018}\cite{choi_xr-based_2022}\cite{zaman2023vicarious}\\
Highlight & \cite{betancourt_exocentric_2022}\cite{xue_enabling_2020}\cite{xu_shared-control_2022}\cite{jang_omnipotent_2021}\cite{xue_er_shamaine_rostar_2020}\cite{brizzi_effects_2018}\cite{szczurek2023multimodal}\\
Haptic & \cite{ponomareva_grasplook_2021}\cite{chen_development_2017}\cite{park_interactive_2018}\cite{rastogi_control_2019}\cite{schwarz_nimbro_2021}\cite{fani_simplifying_2018}\cite{sun_haptic-feedback_2021}\cite{kim2023towards}\cite{xia2023visual}\cite{su_mixed_2021}\cite{yun_immersive_2022}\cite{du_gesture-_2022}\cite{fan2023digital}\cite{fuchino2023t2remoter}\cite{trinitatova2023study}\cite{black2023human}\\
Graphic & \cite{vagvolgyi_scene_2018}\cite{gammieri_coupling_2017}\cite{vempati_virtual_2019}\cite{trejo_user_2018}\cite{gong_real-time_2017}\cite{vu_investigation_2022}\cite{zein_deep_2021}\cite{zinchenko_autonomous_2021}\cite{xue_er_shamaine_rostar_2020}\cite{brizzi_effects_2018}\cite{ihara2023holobots}\cite{sakashita2023vroxy}\cite{zaman2023vicarious}\cite{trinitatova2023study}\\
Avatar & \cite{galambos_design_2015}\cite{takacs_towards_2015}\cite{cousins_development_2017}\cite{honig_mixed_2015}\cite{grzeskowiak_toward_2020}\cite{liu2023multi}\cite{jang_omnipotent_2021}\cite{moniri_human_2016}\cite{choi_xr-based_2022}\cite{ihara2023holobots}\cite{sakashita2023vroxy}\cite{zaman2023vicarious}\\
3D Object & \cite{cruz_ulloa_mixed-reality_2022}\cite{walker_mixed_2021}\cite{mourtzis_augmented_2017}\cite{hernandez_increasing_2020}\cite{walker_robot_2019}\cite{nagy_towards_2019}\cite{mchenry_predictive_2021}\cite{vu_investigation_2022}\cite{mourtzis_augmented_2019}\cite{su_mixed_2021}\cite{zein_deep_2021}\cite{hedayati_improving_2018}\cite{jang_omnipotent_2021}\cite{chacko_augmented_2020}\cite{choi_xr-based_2022}\cite{fuchino2023t2remoter}\cite{trinitatova2023study}\cite{rivera2023toward}\cite{black2023human}\\
No & \cite{chen_enhanced_2022}\cite{xu_design_2022}\cite{wei_multi-view_2021}\cite{stedman_vrtab-map_2022}\cite{lipton_baxters_2017}\cite{wibowo_improving_2021}\cite{kulikovskiy_can_2021}\cite{zhou_intuitive_2020}\cite{stotko_vr_2019}\cite{peppoloni_augmented_2015}\cite{chen_real-time_2020}\cite{yew_immersive_2017}\cite{xie_framework_2022}\cite{ai_real-time_2016}\cite{cichon_digital_2018}\cite{codd-downey_ros_2014}\cite{bai_kinect-based_2018}\cite{peppoloni_immersive_2015}\cite{phan_mixed_2018}\cite{omarali_virtual_2020}\cite{jang_virtual_2021}\cite{wang_human-centered_2019}\cite{mu_design_2021}\cite{kalinov_warevr_2021}\cite{theofilis_panoramic_2016}\cite{zhao_robot_2017}\cite{gai_new_2020}\cite{xu_novel_2018}\cite{le_intuitive_2020}\cite{hormaza_-line_2019}\cite{wang_novel_2017}\\
&\cite{bian_interface_2018}\cite{du_novel_2019}\cite{zhang_extended_2018}\cite{aschenbrenner_collaborative_2018}\cite{nandhini2023teleoperation}\cite{abdulsalam2023vitrob}\cite{nenna2023enhanced}\cite{du2022intelligent}\cite{sanfilippo2022mixed}\\
\bottomrule

\end{tabular}}
\end{table}

\begin{table}[tbh!]
\caption{Systems evaluation methods (DE10)}
\label{Appendix_Table_Evaluation}
\scriptsize
\resizebox{1\linewidth}{!}{ 
\begin{tabular}{p{80pt}p{400pt}}
\toprule
\multicolumn{1}{c}{Category}   & \multicolumn{1}{c}{Citations} \\ \midrule
 N/A & \cite{stedman_vrtab-map_2022}\cite{walker_mixed_2021}\cite{zhou_intuitive_2020}\cite{mourtzis_augmented_2017}\cite{rastogi_control_2019}\cite{yew_immersive_2017}\cite{galambos_design_2015}\cite{gammieri_coupling_2017}\cite{xie_framework_2022}\cite{cichon_digital_2018}\cite{zinchenko_autonomous_2021}\cite{codd-downey_ros_2014}\cite{bai_kinect-based_2018}\cite{takacs_towards_2015}\cite{phan_mixed_2018}\cite{xu_shared-control_2022}\cite{jang_virtual_2021}\cite{wang_human-centered_2019}\cite{mu_design_2021}\cite{theofilis_panoramic_2016}\cite{hernandez_increasing_2020}\cite{zhao_robot_2017}\cite{moniri_human_2016}\cite{xu_novel_2018}\cite{sun_haptic-feedback_2021}\cite{honig_mixed_2015}\cite{hormaza_-line_2019}\cite{xue_er_shamaine_rostar_2020}\cite{wang_novel_2017}\cite{zhang_extended_2018}\\
 &\cite{nagy_towards_2019}\cite{choi_xr-based_2022}\cite{gong_real-time_2017}\cite{mchenry_predictive_2021}\cite{xue_enabling_2020}\cite{aschenbrenner_collaborative_2018}\cite{nandhini2023teleoperation}\cite{steinke2023future}\\
 Time/accuracy of the task & \cite{xu_design_2022}\cite{ponomareva_grasplook_2021}\cite{cruz_ulloa_mixed-reality_2022}\cite{mourtzis_augmented_2019}\cite{zein_deep_2021}\cite{hedayati_improving_2018}\cite{stotko_vr_2019}\cite{fani_simplifying_2018}\cite{omarali_virtual_2020}\cite{trejo_user_2018}\cite{le_intuitive_2020}\cite{walker_robot_2019}\cite{zaman2023vicarious}\cite{kim2023towards}\cite{kanazawa2023considerations}\cite{trinitatova2023study}\cite{meng2023virtual}\cite{nenna2023enhanced}\cite{xia2023visual}\cite{liu2023multi}\cite{black2023human}\cite{wei_multi-view_2021}\cite{vu_investigation_2022}\cite{lipton_baxters_2017}\cite{su_mixed_2021}\cite{wibowo_improving_2021}\cite{vagvolgyi_scene_2018}\cite{kulikovskiy_can_2021}\cite{chen_development_2017}\cite{park_interactive_2018}\cite{peppoloni_augmented_2015}\\
&\cite{wang_virtual_2019}\cite{vempati_virtual_2019}\cite{peppoloni_immersive_2015}\cite{wang_digital_2021}\cite{du_gesture-_2022}\cite{chacko_augmented_2020}\cite{brizzi_effects_2018}\cite{grzeskowiak_toward_2020}\cite{wang_modeling_2019}\cite{du_novel_2019}\cite{fan2023digital}\cite{du2022intelligent}\\
 Questionnaire & \cite{ponomareva_grasplook_2021}\cite{cruz_ulloa_mixed-reality_2022}\cite{mourtzis_augmented_2019}\cite{zein_deep_2021}\cite{hedayati_improving_2018}\cite{stotko_vr_2019}\cite{schwarz_nimbro_2021}\cite{fani_simplifying_2018}\cite{omarali_virtual_2020}\cite{kalinov_warevr_2021}\cite{trejo_user_2018}\cite{le_intuitive_2020}\cite{walker_robot_2019}\cite{ihara2023holobots}\cite{zaman2023vicarious}\cite{kim2023towards}\cite{kanazawa2023considerations}\cite{trinitatova2023study}\cite{meng2023virtual}\cite{nenna2023enhanced}\cite{xia2023visual}\cite{liu2023multi}\cite{rivera2023toward}\\
 Comparison & \cite{xu_design_2022}\cite{yun_immersive_2022}\cite{betancourt_exocentric_2022}\cite{cousins_development_2017}\cite{abdulsalam2023vitrob}\cite{trinitatova2023study}\cite{xia2023visual}\\
 Interview & \cite{hedayati_improving_2018}\cite{walker_robot_2019}\cite{ihara2023holobots}\cite{sanfilippo2022mixed}\\
 AR/VR & \cite{chen_real-time_2020}\cite{black2023human}\\
Other & \cite{chen_real-time_2020}\cite{yun_immersive_2022}\cite{ai_real-time_2016}\cite{gai_new_2020}\\

\bottomrule

\end{tabular}}
\end{table}

\begin{table}[tbh!]
\caption{Whether the system supports multiplayer/multi-bot (only included supported categories, DE12)}
\label{Appendix_Table_Multi}
\scriptsize
\resizebox{1\linewidth}{!}{ 
\begin{tabular}{p{100pt}p{380pt}}
\toprule
\multicolumn{1}{c}{Category}   & \multicolumn{1}{c}{Citations} \\ \midrule
Multi-user - One robot & \cite{galambos_design_2015}\\
One user - Multi-robot & \cite{jang_omnipotent_2021}\cite{gong_real-time_2017}\\
Multi-user - Multi-robot & \cite{phan_mixed_2018}\cite{honig_mixed_2015}\cite{aschenbrenner_collaborative_2018}\\
One-user - One-user+One robot & \cite{mourtzis_augmented_2019}\cite{mourtzis_augmented_2017}\cite{schwarz_nimbro_2021}\cite{moniri_human_2016}\cite{sakashita2023vroxy}\cite{zaman2023vicarious}\cite{szczurek2023multimodal}\cite{fuchino2023t2remoter}\cite{liu2023multi}\cite{black2023human}\\
One-user - Multi-user+One robot & \cite{walker_mixed_2021}\\

\multicolumn{2}{l}{\color[HTML]{82C1DA}Included articles that are not listed on the table i.e. do not support multiplayer/multi-bot operations.}\\
\bottomrule

\end{tabular}}
\end{table}

\end{document}